\documentclass[submitting]{nst}

\usepackage{subfigure,dcolumn}
\usepackage{epstopdf}
\usepackage{mhchem}
\usepackage{upgreek}
\usepackage{mathrsfs}
\usepackage{longtable}
\usepackage{bm}
\usepackage{tabularx}
\usepackage{array}
\usepackage{hyperref}
\usepackage{multirow}
\usepackage{lineno}

\begin{document}
\nolinenumbers
\title{Benchmarking nuclear energy density functionals with new mass data}
\thanks{Supported by the National Natural Science Foundation of China (Grants No. 12265012 and No. 12305125), Guizhou Provincial Science and Technology Projects (Grant No. ZK[2022]203), PhD fund of Guizhou Minzu University (Grant No. GZMUZK[2024]QD76), the National Key Laboratory of Neutron Science and Technology (Grant No. NST202401016), and the Sichuan Science and Technology Program (Grant No. 2024NSFSC1356). Fruitful discussions with members of the DRHBc Mass Table Collaboration are greatly appreciated.}

\author{Xiaoying Qu}
\affiliation{School of Physics and Mechatronic Engineering, Guizhou Minzu University, Guiyang 550025, Guizhou China}

\author{Kangmin Chen}
\affiliation{School of Physics and Mechatronic Engineering, Guizhou Minzu University, Guiyang 550025, Guizhou China}

\author{Cong Pan}
\affiliation{Department of Physics, Anhui Normal University, Wuhu 241000, Anhui, China}

\author{Yangyang Yu}
\affiliation{School of Physics and Mechatronic Engineering, Guizhou Minzu University, Guiyang 550025, Guizhou China}

\author{Kaiyuan Zhang}
\email[Corresponding author, ]{zhangky@caep.cn}
\affiliation{National Key Laboratory of Neutron Science and Technology, Institute of Nuclear Physics and Chemistry, China Academy of Engineering Physics, Mianyang 621900, Sichuan, China}

\begin{abstract}
Nuclear masses play a crucial role in both nuclear physics and astrophysics, driving sustained efforts toward their precise experimental determination and reliable theoretical prediction.
In this work, we compile the newly measured masses for 296 nuclides from 40 references published between 2021 and 2024, subsequent to the release of the latest Atomic Mass Evaluation.
These data are used to benchmark the performance of several relativistic and non-relativistic density functionals, including PC-PK1, TMA, SLy4, SV-min, UNEDF1, and the recently proposed PC-L3R.
Results for PC-PK1 and PC-L3R are obtained using the state-of-the-art deformed relativistic Hartree-Bogoliubov theory in continuum (DRHBc), while the others are adopted from existing literature.
It is found that the DRHBc calculations with PC-PK1 and PC-L3R achieve an accuracy better than 1.5 MeV, outperforming the other functionals, which all exhibit root-mean-square deviations exceeding 2 MeV.
The odd-even effects and isospin dependence in these theoretical descriptions are examined.
The PC-PK1 and PC-L3R descriptions are qualitatively similar, both exhibiting robust isospin dependence along isotopic chains.
Finally, a quantitative comparison between the PC-PK1 and PC-L3R results is presented, with their largest discrepancies analyzed in terms of potential energy curves from the constrained DRHBc calculations.
\end{abstract}


\maketitle
\nolinenumbers
\section{Introduction}\label{section1}

Nuclear mass or binding energy reflects the complex nuclear forces binding protons and neutrons together within a nucleus \cite{Wienholtz2013Nature}.
This fundamental quantity not only underlies the nuclear stability \cite{Ramirez2012Science337} but also critically influences astrophysical phenomena--from the nuclear reactions in stellar interiors \cite{Bethe1939PR55} to the nucleosynthesis processes responsible for elemental production in the universe \cite{Burbidge1957RMP}.
Consequently, precise determinations of nuclear masses are indispensable for advancing our understanding of nuclear structure \cite{Lunney2003RMP} and have significant implications for nuclear astrophysics \cite{Aprahamian2005PPNP,Schatz2013IJMS,Mumpower2016PPNP}.
Improved experimental precision and theoretical accuracy in nuclear mass evaluations thus not only deepen insights into the fundamental research in nuclear physics \cite{Gao2023NST,Yang2025NST} but also foster progress in nuclear energy applications via both fusion and fission.

Global investments in rare isotope beam facilities--including the Heavy Ion Research Facility in Lanzhou (HIRFL) \cite{Yuan2020JPCS} and the High Intensity heavy-ion Accelerator Facility (HIAF) at Huizhou \cite{ZhouXH2022AAPPS}, China, the Facility for Rare Isotope Beams (FRIB) in the USA \cite{Davide2022Nature201}, the Radioactive Isotope Beam Factory (RIBF) at RIKEN, Japan \cite{Sakurai2018FP}, the Facility for Antiproton and Ion Research (FAIR) in Germany \cite{Durante2019PS}, the Rare isotope Accelerator complex for ON-line experiments (RAON) in Korea \cite{Hong2023JPCS},
and the Isotope Separator and Accelerator in Canada (ISAC) \cite{Ball2011JPG}--have substantially advanced the production, identification, and investigation of nuclides far from the valley of stability.
To date, experimental efforts have led to the identification of over 3300 nuclides \cite{Thoennessen2016book}, with mass measurements available for approximately 2500 of these \cite{Kondev2021CPC,Huang2021CPC,WangM2021CPC,Shi2024NST}.
In contrast, theoretical models predict the existence of approximately 7000--10000 nuclides \cite{Erler2012Nature,Xia2018ADNDT}.
Given that the proton dripline has been established for isotopes with proton numbers $Z \gtrsim 90$ \cite{Zhang2019PRL}, while the neutron dripline has been delineated only up to $Z=10$ \cite{Ahn2019PRL}, it is anticipated that most unknown neutron-rich nuclei will remain experimentally inaccessible in the foreseeable future.
Therefore, there is an urgent need for reliable theoretical predictions of nuclear masses.

Extensive efforts have been devoted to both reproducing measured nuclear masses and predicting those yet uncharted.
Macroscopic-microscopic approaches exemplified by the finite-range droplet model (FRDM) \cite{Moller2016ADNDT} and the Weizs\"acker-Skyrme (WS) model \cite{Wang2014PLB,Wan2025NST} have achieved impressive accuracy in describing existing mass data; however, microscopic theories are widely accepted as offering superior predictive capabilities \cite{ZhangKY2021PRC,He2024PRC}.
In this context, density functional theory has emerged as a powerful framework for a unified description of nearly all nuclides across the nuclear chart \cite{Hirata1997NPA,Lalazissis1999ADNDT,Geng2005Prog.Theor.Phys.785,Goriely2009PRL152503,Goriely2009PRL24,Afanasjev2013PLB,Agbemava2014PRC,Lu2015Phys.Rev.C27304,YangYL2021PRC}.
Its relativistic extension--the covariant density functional theory (CDFT) \cite{Meng2016book}--has proven exceptionally successful in describing a variety of nuclear phenomena in both ground and excited states \cite{Meng2016book,Ring1996PPNP,Vretenar2005PR,Meng2006PPNP,Niksic2011PPNP,Meng2013FP,Meng2015JPG,Zhou2016PS,ShenSH2019PPNP,Meng2021AAPPS}.
This success is largely attributable to the inherent advantages of CDFT, including the automatic incorporation of spin-orbit coupling \cite{Kucharek1991ZPA,RenZX2020PRC}, the natural explanation of pseudospin symmetry in the nucleon spectrum \cite{Ginocchio1997PRL,MengJ1998PRC,Liang2015PR} and spin symmetry in the antinucleon spectrum \cite{Zhou2003PRL,HeXT2006EPJA,Liang2015PR}, as well as its self-consistent treatment of nuclear magnetism \cite{Koepf1989NPA,Koepf1990NPA}.

Within the framework of CDFT, the pairing correlations and continuum effects are taken into account self-consistently in the relativistic continuum Hartree-Bogoliubov (RCHB) theory \cite{Meng1996PRL,Meng1998NPA}, making it capable of describing both stable and exotic nuclei \cite{Meng1996PRL,Meng1998PRL,ZhangSQ2002CPL,Meng2002PLB,ZhangW2005NPA,Lim2016PRC,Kuang2023EPJA}.
A pioneering application of the RCHB theory is the construction of the first relativistic nuclear mass table incorporating continuum effects, in which the existence of 9035 bound nuclei with $8\le Z\le 120$ is predicted \cite{Xia2018ADNDT}.
Notably, the inclusion of continuum effects is found essential in extending the neutron dripline towards the more neutron-rich region.
Nonetheless, the accuracy of the RCHB mass table in reproducing experimental data is limited, due to the assumption of spherical symmetry within the theoretical framework.

It is thus natural to propose an upgraded mass table that incorporates not only the continuum effects but also nuclear deformation degrees of freedom.
This can be realized by employing the deformed extension of the RCHB theory, the deformed relativistic Hartree-Bogoliubov theory in continuum (DRHBc) \cite{Zhou2010PRC,LiLL2012PRC,Li2012CPL,Chen2014PRC}.
The axial deformation, pairing correlations, and continuum effects are taken into account microscopically and self-consistently in the DRHBc theory, which lays an important foundation for its great success \cite{Meng2015JPG,Sun2024NPR,Zhang2024NPR}.
In pursuit of a high-precision mass table \cite{Zhang2021CSB}, the point-coupling version of the DRHBc theory has been developed \cite{ZhangKY2020PRC102,Pan2022PRC} to combine with the density functional PC-PK1 \cite{Zhao2010Phys.Rev.C54319}, which is probably the most successful density functional for describing nuclear masses \cite{Zhao2012Phys.Rev.C64324,Meng2013FP,Lu2015Phys.Rev.C27304}.
The DRHBc mass table project, now in progress for over six years, has successfully completed the sectors for even-even \cite{Zhang2022ADNDT} and even-$Z$ \cite{Guo2024ADNDT} nuclei.
Impressively, the root-mean-square (RMS) deviation of the DRHBc calculated masses from the latest Atomic Mass Evaluation (AME2020) data is approximately 1.5 MeV, positioning it among the most accurate density-functional descriptions for nuclear masses.
Moreover, lots of relevant studies on
halo phenomena \cite{Sun2018PLB,ZhangKY2019PRC,SunXX2020NPA,YangZH2021PRL,SunXX2021PRC,Zhong2022SCP,ZhangKY2023PLB,ZhangKY2023PRC107,Pan2024PLB,AnJL2024PLB,WangLY2024EPJA},
nuclear charge radii \cite{ZhangXY2023PRC,Mun2023PLB,Pan2025arXiv},
shape evolution \cite{GuoP2023PRC,MunMH2024PRC024310,Choi2022PRC,Kim2022PRC},
shell structure \cite{ZhengRY2024CPC,ZhangYX2024PRC,LiuWJ2024Particle,DuP2024Particle,PanC2025Particle,LiangW2025Particle,HuangZD2025PRC},
decay properties \cite{XiaoY2023PLB,Choi2024PRC,Lu2024PLB,MunMH2025Part82},
and other topics \cite{PanC2021PRC,HeXT2021CPC,HeXT2024PRC,MunMH2024PRC014314,ZhangW2024CPC,WangSB2024Particle} based on the DRHBc mass table underscore its value as a resource that extends far beyond a mere data repository \cite{ZhangKY2025AAPPS}.

In this work, inspired by the recent progress in nuclear mass measurements that provide new data beyond AME2020 or reduce the uncertainties of existing data, we further examine the predictive power of the DRHBc mass table using the new mass data.
On the theoretical side, a new point-coupling density functional, PC-L3R, has recently been proposed, whose performance would be even better than PC-PK1 in describing masses of spherical nuclei \cite{Liu2023PLB}.
Our second motivation is to test the accuracy of PC-L3R in describing masses of deformed nuclei when combined with the DRHBc theory.
This article is structured as follows.
The point-coupling DRHBc theory, the relativistic density functionals PC-PK1 and PC-L3R, and the numerical details are introduced in section \ref{section2}.
The DRHBc descriptions with PC-PK1 and PC-L3R for the new masses are presented and compared with those from other density functionals in section \ref{section3}.
Finally, a summary is given in section \ref{section4}.

\section{Theoretical framework}\label{section2}

The point-coupling density functional theory starts from the Lagrangian density,
\begin{equation}
\begin{split} \mathcal{L}=&\bar\psi(i\gamma_{\mu}\partial^{\mu}-M)\psi
-\frac{1}{2}\alpha_S(\bar\psi\psi)(\bar\psi\psi)
-\frac{1}{2}\alpha_V(\bar\psi\gamma_{\mu}\psi)(\bar\psi\gamma^{\mu}\psi)\\
&-\frac{1}{2}\alpha_{TV}(\bar\psi\vec\tau\gamma_{\mu}\psi)(\bar\psi\vec{\tau}\gamma^{\mu}\psi)
-\frac{1}{2}\alpha_{TS}(\bar\psi\vec\tau\psi)(\bar\psi\vec\tau\psi)\\
&-\frac{1}{3}\beta_S(\bar\psi\psi)^3-\frac{1}{4}\gamma_S(\bar\psi\psi)^4
-\frac{1}{4}\gamma_V[(\bar\psi\gamma_{\mu}\psi)(\bar\psi\gamma^{\mu}\psi)]^2\\
&-\frac{1}{2}\delta_S\partial_{\nu}(\bar\psi\psi)\partial^{\nu}(\bar\psi\psi)
-\frac{1}{2}\delta_V\partial_{\nu}(\bar\psi\gamma_{\mu}\psi)\partial^{\nu}(\bar\psi\gamma^{\mu}\psi)\\
&-\frac{1}{2}\delta_{TV}\partial_{\nu}(\bar\psi\vec\tau\gamma_{\mu}\psi)\partial^{\nu}(\bar\psi\vec\tau\gamma^{\mu}\psi)\\
&-\frac{1}{2}\delta_{TS}\partial_{\nu}(\bar\psi\vec\tau\psi)\partial^{\nu}(\bar\psi\vec\tau\psi)\\
&-\frac{1}{4}F^{\mu\nu}F_{\mu\nu}-e\bar\psi\gamma^{\mu}\frac{1-\tau_3}{2}A_{\mu}\psi,
\end{split}\label{Lagrangian}
\end{equation}
where $M$ is the nucleon mass, $e$ the charge unit, and $A_{\mu}$ and $F_{\mu\nu}$ the four-vector potential and field strength tensor of the electromagnetic field, respectively.
With the subscripts $S$, $V$, and $T$ respectively standing for scalar, vector, and isovector, nine coupling constants, $\alpha_{S}$, $\alpha_V$, $\alpha_{TV}$, $\beta_S$, $\gamma_S$, $\gamma_V$, $\delta_S$, $\delta_V$, and $\delta_{TV}$, in the Lagrangian density of PC-PK1 and PC-L3R are listed in Table \ref{parameters}.
As the isovector-scalar channels involving $\alpha_{TS}$ and $\delta_{TS}$ terms were found less helpful to improve the description of nuclear ground-state properties \cite{Burvenich2002PRC}, they are not included in PC-PK1 and PC-L3R.
\begin{table}[h!]
\centering
\caption{Coupling constants of the relativistic density functionals PC-PK1 \cite{Zhao2010Phys.Rev.C54319} and PC-L3R \cite{Liu2023PLB}.}
\label{parameters}
\begin{tabular}{ccccc}
\hline\hline
Coupling constant & PC-PK1  & PC-L3R  & Unit   \\\hline
$\alpha_S$    &  $-3.96291\times 10^{-4}$  & $-3.99289\times 10^{-4}$ & MeV$^{-2}$ \\
$\beta_S$     &  $8.6653\times 10^{-11}$    & $8.65504\times 10^{-11}$ & MeV$^{-5}$\\
$\gamma_S$    &  $-3.80724\times 10^{-17}$  & $-3.83950\times 10^{-17}$& MeV$^{-8}$ \\
$\delta_S$    &  $-1.09108\times 10^{-10}$  & $-1.20749\times 10^{-10}$
& MeV$^{-4}$ \\
$\alpha_V$    &  $2.6904\times 10^{-4}$     & $2.71991\times 10^{-4}$ & MeV$^{-2}$  \\
$\gamma_V$    &  $-3.64219\times 10^{-18}$  & $-3.72107\times 10^{-18}$& MeV$^{-8}$   \\
$\delta_V$    &  $-4.32619\times 10^{-10}$  & $-4.26653\times 10^{-10}$& MeV$^{-4}$  \\
$\delta_{TV}$ &  $2.95018\times 10^{-5}$    & $2.96688\times 10^{-5}$& MeV$^{-2}$  \\
$\delta_{TV}$ &  $-4.11112\times 10^{-10}$  & $-4.65682\times 10^{-10}$& MeV$^{-4}$  \\
\hline \hline
\end{tabular}
\end{table}

Starting from the Lagrangian density \eqref{Lagrangian}, the Hamiltonian can be derived via the quantization the Dirac spinor field in the Bogoliubov quasi-particle space, and the energy functional can be constructed as its expectation with respect to the Bogoliubov ground state.
The relativistic Hartree-Bogoliubov equation obtained by performing the variation of the energy density functional with respect to the generalized density matrix and neglecting the exchange terms reads
\begin{equation}\label{RHBequ}
\left(
\begin{array}{cc}
  \hat h_D-\lambda   &  \hat\Delta  \\
  -\hat\Delta^*            &  -\hat h^*_D+\lambda
\end{array}
 \right)
\left(
\begin{array}{c}
U_k\\
V_k
\end{array}
\right)
=E_k\left(
\begin{array}{c}
U_k\\
V_k
\end{array}
\right),
\end{equation}
where $\hat h_D$ is the Dirac Hamiltonian, $\hat\Delta$ is the pairing field, $\lambda$ is the Fermi surface, $E_{k}$ is the quasiparticle energy, and $U_{k}$ and $V_k$ are quasiparticle wave functions.
The Dirac Hamiltonian in coordinate space is
\begin{equation}
h_{D}(\bm r)=\bm{\alpha} \cdot \bm p+V(\bm r)+\beta [M+S(\bm r)],
\end{equation}
with the scalar $S(\bm r)$ and vector $V(\bm r)$ potentials,
\begin{equation}
\begin{split}
S(\bm r)=&\alpha_S\rho_S+\beta_S\rho^2_{S}+\gamma_S\rho^3_{S}+\delta_S\Delta\rho_S,\\
V(\bm r)=&
\alpha_V\rho_V+\gamma_V\rho_V^3+\delta_V\Delta\rho_V+eA^0+\alpha_{TV}\tau_3\rho_3\\
&+\delta_{TV}\tau_3\Delta\rho_3,
\end{split}
\end{equation}
constructed by various densities,
\begin{equation}
\begin{split}
\rho_S(\bm r)&=\sum_{k>0}V_k^{\dagger}(\bm r)\gamma_0 V_k(\bm r),\\
\rho_V(\bm r)&=\sum_{k>0}V_k^{\dagger}(\bm r)V_k(\bm r),\\
\rho_3(\bm r)&=\sum_{k>0}V_k^{\dagger}(\bm r)\tau_3 V_k(\bm r),
\end{split}
\end{equation}
where the summation index $k$ loops through the positive-energy quasiparticle states in the Fermi sea.
Neglecting for simplicity spin and isospin degrees of freedom, the pairing potential reads
\begin{equation}\label{pairpoten}
\Delta(\bm r_1,\bm r_2)=V^{pp}(\bm r_1,\bm r_2)\kappa(\bm r_1,\bm r_2),
\end{equation}
where $V^{pp}$ is the pairing interaction and $\kappa$ the pairing tensor.
A density-dependent interaction of zero range is adopted in the present DRHBc theory,
\begin{equation}\label{pairforce}
V^{pp}(\bm r_1,\bm r_2)=V_0\frac{1}{2}(1-P^{\sigma})\delta(\bm r_1-\bm r_2)
\left(
1-\frac{\rho(\bm r_1)}{\rho_{\rm sat}}
\right),
\end{equation}
with $V_0$ being the pairing strength, $\rho_{\rm sat}$ as the saturation density of nuclear matter, and $\frac{1}{2}(1-P^{\sigma})$ projecting onto the spin-zero $S=0$ component.
Assuming axial symmetry and spatial reflection symmetry, one can expand nuclear densities and potentials in terms of even-order Legendre polynomials \cite{Price1987PRC},
\begin{equation}\label{flambda}
    f(\bm r)=\sum_{l}f_l(r)P_l(\cos\theta),~~l=0,2,4,\cdots,
\end{equation}
where the $l$th-order radial component is calculated by
\begin{equation}
f_l(r)=\frac{2l+1}{4\pi}\int d\Omega f(\bm r)P_l(\cos\theta).
\end{equation}

After obtaining the self-consistent solution of the RHB equation \eqref{RHBequ}, quantities including the total energy, the rms radius, and the deformation parameter can be calculated from nuclear densities.
The total energy is \cite{Meng1998NPA}
\begin{equation}
\begin{split}
E_{\rm RHB}&=\sum_{k>0}(\lambda-E_k)v_k^2-E_{\rm pair}\\
&-\int d^3\bm r\left(
\frac{1}{2}\alpha_S\rho^2_S+\frac{1}{2}\alpha_V\rho_V^2+\frac{1}{2}\alpha_{TV}\rho_3^2\right.\\
&+\frac{2}{3}\beta_S\rho_S^3+\frac{3}{4}\gamma_S\rho^4_S+\frac{3}{4}\gamma_V\rho_V^4+\frac{1}{2}\delta_S\rho_S\Delta\rho_S\\
&\left. +\frac{1}{2}\delta_V\rho_V\Delta\rho_V+\frac{1}{2}\delta_{TV}\rho_3\Delta\rho_3+\frac{1}{2}\rho_peA^0
\right) \\
&+ E_{\text{c.m.}} + E_{\text{rot}},
\end{split}
\end{equation}
where
\begin{equation}
v_k^2=\int d^3\bm r V_k^{\dagger}(\bm r)V_k(\bm r),
\end{equation}
and the pairing energy
\begin{equation}
E_{\rm pair} = -\frac{1}{2}\int d^3\bm r\kappa(\bm r) \Delta (\bm r).
\end{equation}
The center-of-mass (c.m.) correction energy is calculated as
\begin{equation}
E_{\rm c.m.}=-\frac{1}{2MA}\langle \hat P^2 \rangle,
\end{equation}
with $A$ the mass number and $\hat P=\sum^{A}_{i}\hat p_i$ the total momentum in the c.m. frame.
The rotational correction energy for a deformed nucleus from the cranking approximation reads
\begin{equation} \label{erot}
E_{\rm rot}=-\frac{1}{2\mathscr J}\langle \hat J^2 \rangle,
\end{equation}
where $\hat J=\sum^A_i\hat j_i$ is the total angular momentum and $\mathscr J$ is the moment of inertia that can be estimated via the Inglis-Belyaev formula \cite{Peter1980}.
The RMS radius is calculated as
\begin{equation}
    R_{\rm m}=\langle r^2\rangle ^{1/2}=\sqrt{\frac{\int d^3 \bm r [r^2 \rho_V(\bm r)]}{N}},
\end{equation}
where $\rho_V$ is the vector density, and $N$ denotes the corresponding particle number.
The quadrupole deformation parameter is calculated as
\begin{equation}
\beta_2=\frac{4\pi\langle r^2Y_{20}(\Omega)\rangle}{3N\langle r^2 \rangle},
\end{equation}
with $Y$ being the spherical harmonic function.

The calculations in this work are performed using the same numerical details as those in constructing the DRHBc mass table \cite{ZhangKY2020PRC102,Pan2022PRC,Zhang2022ADNDT}.
Specifically, the pairing strength $V_0=-325$ MeV fm$^3$, the saturation density $\rho_{\rm sat} = 0.152$ fm$^{-3}$, and the pairing window is taken as 100 MeV.
The Dirac Woods-Saxon basis space is determined by an energy cutoff of $E_{\rm cut}=300$ MeV and an angular momentum cutoff of $J_{\rm cut}=\frac{23}{2}~\hbar$.
The Legendre expansion truncations in Eq. \eqref{flambda} are chosen as $l_{\rm max} = 6$ and 8 for the nuclei with $8 \leq Z \leq 70$ and $71 \leq Z \leq 100$, respectively.
The blocking effects in odd-mass and odd-odd nuclei are included via the equal filling approximation \cite{Perez2008PRC,Li2012CPL,Pan2022PRC}.

\section{Results and discussion}\label{section3}

\begin{figure}[!htb]
\centering
\includegraphics[width=3.2in]{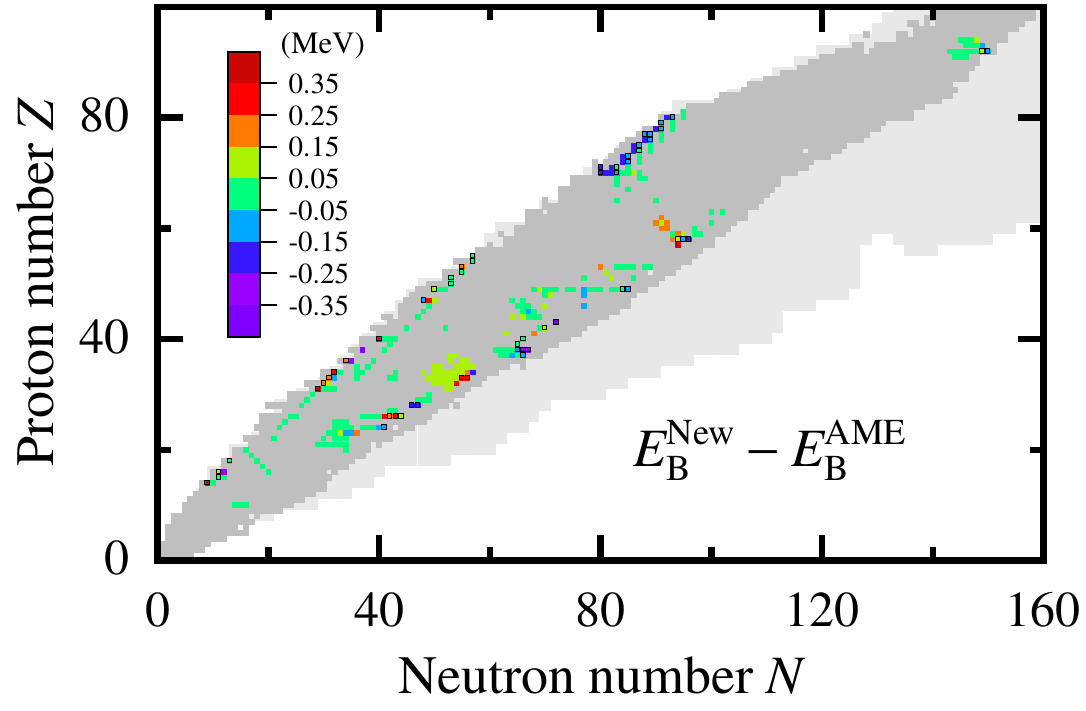}
\caption{The differences between the newly measured masses for 296 nuclides and the corresponding values in AME2020 \cite{WangM2021CPC} scaled by colors.
The black squares represent nuclides for which AME2020 has only empirical mass values.
The dark gray region shows the nuclides observed experimentally \cite{NNDC}, and the light gray region shows the predicted nuclear landscape by the DRHBc mass table \cite{Zhang2022ADNDT}.}
\label{fig1}
\end{figure}

We have collected the newly measured masses of 296 nuclides from 40 references \cite{jacobs2023improved,YUY2024PRL133,Puentes106PRC,Yandow2023PRC,Surbrook2021PRC,Porter2022PRC106,Lebit2021PRL,WangM2022PRC,Iimura2023PRL,Silwal2022PLB,
Porter2022PRC105,Canete2021PRC,giraud2022mass,wang2023mass,PhysRevC.104.065803,XianW2024PRC,xing2023isochronous,Horana2022PRC,zhou2023Mass,PhysRevC.103.034319,
Mukul2021PRC,Hukkanen2024PLB,Hamaker2021NaturePhys,Hou2023PRC,WangKL2024PRC,PorterPRC110,LiHF2022PRL,GeZh2024PRL,JariesPRC2023,mougeot2021mass,Nesterenko2023PRC,
Izzo2021PRC,Hoff2023PRL,Valverde2024PLB,Beliuskina2024PRC,Kimura2024PRC,Orford2022PRC,lykiardopoulou2023exploring,Beck2021PRL,Niwase2023PRL},
published between 2021 and 2024 (subsequent to the release of AME2020), and summarized them in Table \ref{mass-table}.
The sources and measurement methods of the new experimental data are tabulated in Table \ref{exp-method}.
The corresponding mass values in AME2020, which contains both 241 experimentally measured values and 55 extrapolated empirical values (labeled \#), are listed in Table \ref{mass-table} for comparison.
The differences between the new data and the AME2020 values are also given in Table \ref{mass-table} and scaled by colors in Fig. \ref{fig1}.
The nuclides with only empirical values in AME2020 are highlighted by black squares in Fig. \ref{fig1}.
It can be found that among these 296 mass data, 247 in AME2020 are consistent with new measurements with their deviations smaller than 0.15 MeV.
The rms deviation between the new and AME2020 experimental data is $\sigma=0.0984$ MeV, and after including the empirical values in AME2020, $\sigma$ becomes 0.1178 MeV.

\begin{figure}[!htb]
\centering
\includegraphics[width=3.2in]{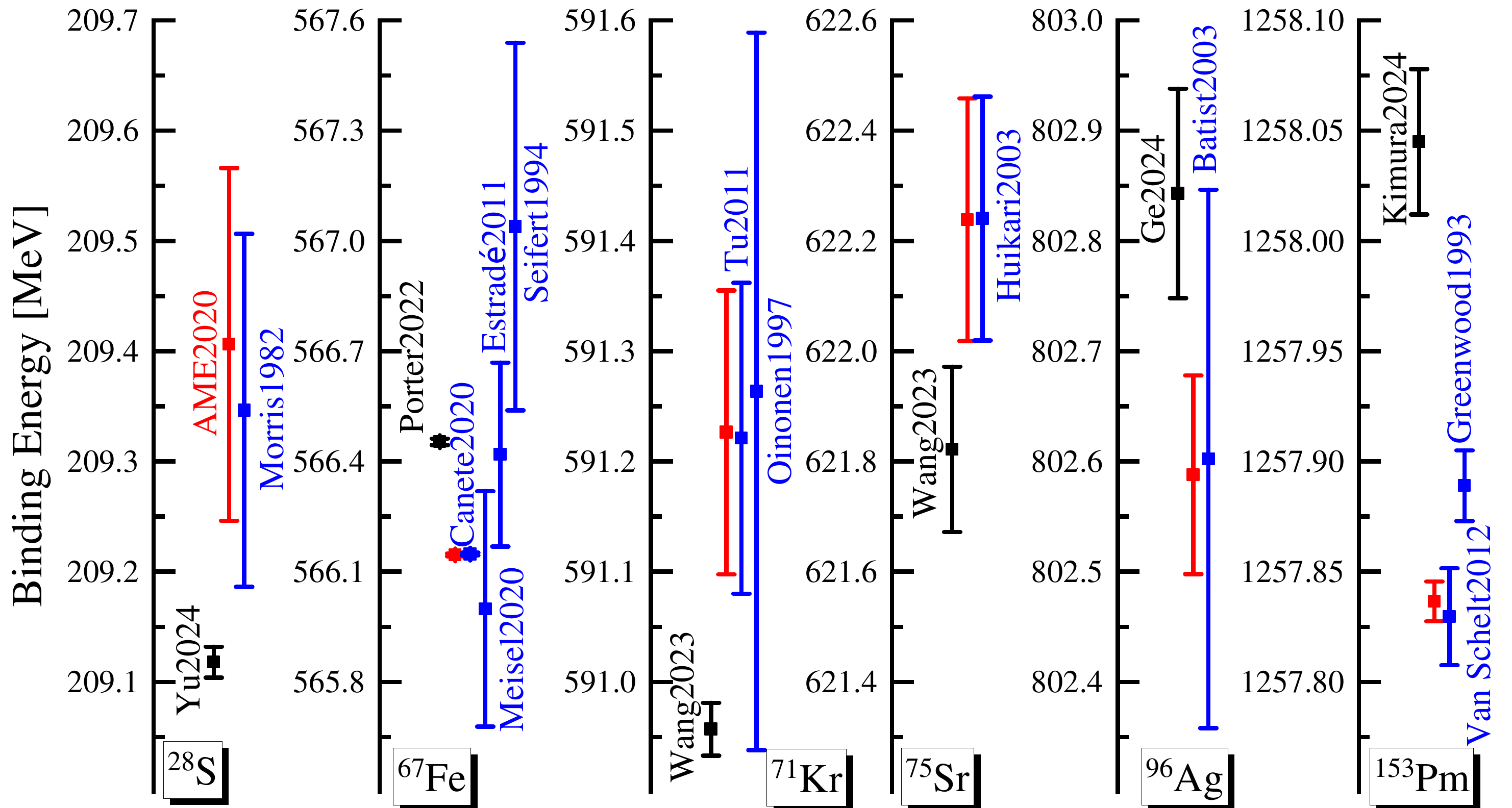}
\caption{The newly measured binding energies of $^{28}$S, $^{67}$Fe, $^{71}$Kr, $^{75}$Sr, $^{96}$Ag, and $^{153}$Pm, in comparison with the AME2020 data and the previous data referenced in the AME.}
\label{fig2}
\end{figure}

For the deviations between the new and AME2020 experimental data, most of them lie within the experimental uncertainties.
Nevertheless, even after considering experimental uncertainties, disagreement still arises for 73 nuclides, which are labeled in red and bold in Table \ref{mass-table}.
Among these 73 data, the smallest difference, $0.00055$ MeV, occurs between the upper bound of the new data and the lower bound of the AME2020 data for $^{111}$Ag, while the largest one, $0.29516$ MeV, occurs between the lower bound of the new data and the upper bound of the AME2020 data for $^{67}$Fe.
It is important to note that newly measured masses are not necessarily more accurate than previous evaluations. 
Systematic biases or experimental uncertainties may affect the measured values, depending on the specific setup and techniques employed. 
Notably, for 23 nuclides, the central values of the newly reported masses deviate from those in AME2020 by more than 200 keV. 
Among the 23 nuclides exhibiting discrepancies exceeding 200 keV, 12 cases ($^{23}$Si, $^{74}$Ni, $^{86}$Ge, $^{89}$As, $^{91}$Se, $^{70}$Kr, $^{104}$Sr, $^{105}$Sr, $^{109}$Nb, $^{72}$Tc, $^{151}$La, and $^{151}$Yb) show overlapping uncertainty ranges between the new measurements and AME2020 values, indicating potential consistency within experimental uncertainties. 
For 5 nuclides ($^{69}$Fe, $^{60}$Ga, $^{88}$As, $^{66}$Se, and $^{80}$Zr), while the central value discrepancies also exceed 200 keV and uncertainty ranges do not overlap, the AME2020 masses are extrapolated values. 
Although such empirical estimates, based on trends in the mass surface and available experimental constraints, are often validated by subsequent measurements, deviations from true mass values may arise, e.g., for nuclides exhibiting abrupt changes in shell structure. 
Finally, for 6 nuclides ($^{28}$S, $^{67}$Fe, $^{71}$Kr, $^{75}$Sr, $^{96}$Ag, and $^{153}$Pm), the central values differ by more than 200 keV, the uncertainty ranges do not overlap, and both the new and AME2020 masses are based on experimental data.
These cases are compared in Fig.~\ref{fig2}, alongside the earlier measurements referenced in AME2020. 
The observed discrepancies may arise from differences in measurement techniques. 
For $^{28}$S, its mass was derived from an indirect measurement in 1982 \cite{Morris1982PRC}, whereas the recent result is from a direct measurement using the B$\rho$-defined isochronous mass spectrometry (B$\rho$-IMS) in 2024 \cite{YUY2024PRL133}. 
For $^{67}$Fe, AME2020 mainly adopted the ion trap data from 2020 \cite{Canete2020PRC}, which fall within the uncertainty range of contemporaneous B$\rho$-time-of-flight (B$\rho$-TOF) measurements \cite{Meisel2020PRC}.
However, the 2022 result using a multiple-reflection TOF mass spectrometer (MR-TOF-MS) \cite{Porter2022PRC105} supports earlier data from the TOF isochronous spectrometer from 2011 \cite{Estrade2011PRL} rather than that from 1994 \cite{Seifert1994ZPA}. 
For $^{71}$Kr, the AME2020 value is consistent with storage-ring IMS data \cite{Tu2011PRL}. 
While the new B$\rho$-IMS result from 2023 \cite{wang2023mass} shows deviations, it remains largely within the uncertainty range of the results inferred from electron-capture decay energy $Q_{\mathrm{EC}}$ measurements \cite{Oinonen1997PRC}. 
For $^{75}$Sr and $^{96}$Ag, the AME2020 values are consistent with earlier estimates based on $Q_{\mathrm{EC}}$ \cite{Huikari2003EPJA, Batist2003NPA}, but deviate from recent measurements obtained via B$\rho$-IMS \cite{wang2023mass,Zhou2024NST} and ion trap techniques \cite{GeZh2024PRL}, respectively. 
For $^{153}$Pm, AME2020 primarily adopts the ion trap data from 2012 \cite{Van2012PRC}.
Nonetheless, discrepancies are observed among the 2012 results, earlier estimates based on $\beta^-$ decay energy from 1993 \cite{Green1993NIMPRA}, and the latest measurement from MR-TOF-MS in 2024 \cite{Kimura2024PRC}.
Such discrepancies necessitate careful data evaluation and/or even further measurements.
In this study, we adopt for convenience the newly measured masses in the following examination of theoretical descriptions.

\begin{figure*}[htbp]
\centering
\includegraphics[width=5 in]{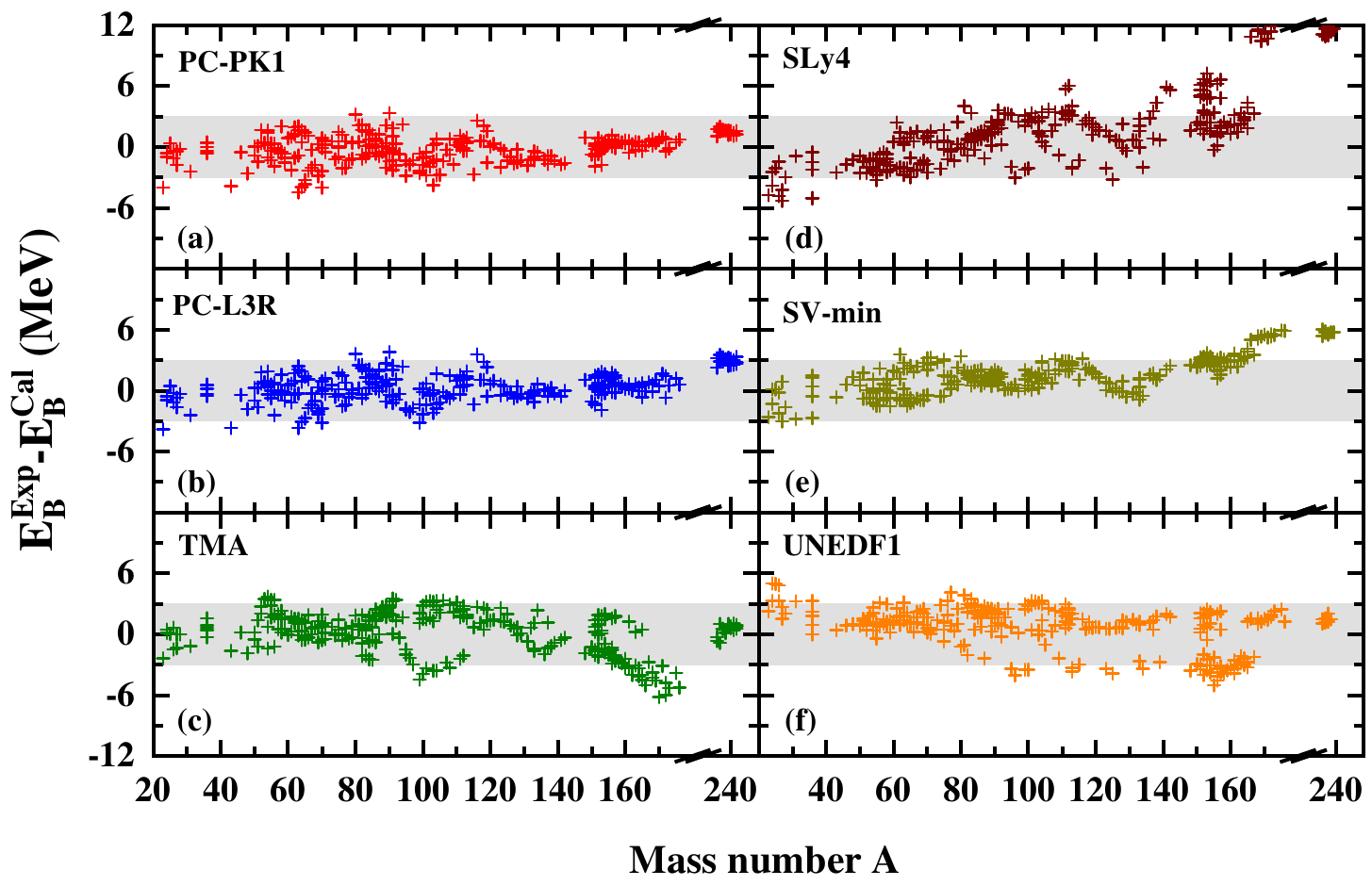}
\caption{Mass differences between new data listed in Table \ref{mass-table} and the values from DRHBc calculations with density functionals PC-PK1 (a), PC-L3R (b), RMF+BCS calculations with density functional TMA \cite{Geng2005Prog.Theor.Phys.785} (c), as well as Skyrme HFB calculations \cite{Massexp} with density functionals SLy4 (d), SV-min (e) and UNEDF1 (f).}
\label{fig3}
\end{figure*}

The DRHBc calculations for the 296 nuclides have been performed with density functionals PC-PK1 \cite{Zhao2010Phys.Rev.C54319} and PC-L3R \cite{Liu2023PLB}, and the deviations of the resulting nuclear masses from the experimental data are plotted respectively in Figs.~\ref{fig3}(a) and \ref{fig3}(b).
For comparison, we also exhibit the mass differences between the new data and the results from the relativistic mean-field plus Bardeen-Cooper-Schrieffer (RMF+BCS) calculations with TMA \cite{Geng2005Prog.Theor.Phys.785} and the non-relativistic Skyrme Hartree-Fock-Bogoliubov (HFB) calculations \cite{Massexp} with SLy4 \cite{Chabanat1998NPA}, SV-min \cite{Klupfel2009PRC}, and UNEDF1 \cite{Kortelainen2012PRC}, respectively, in Figs.~\ref{fig3}(c)--(f).
It can be found that both the DRHBc calculations with PC-PK1 and PC-L3R reproduce fairly well the data within a deviation of 3 MeV, despite a few exceptions.
The RMF+BCS calculations with TMA can achieve a similar level of accuracy for the nuclides with $A<150$, but for heavier nuclides an overestimation up to 6 MeV arises.
In contrast, the Skyrme HFB calculations with SLy4 significantly underestimate the data in the heavy mass region, with the deviations for several nuclides above 10 MeV.
The results from Skyrme HFB calculations with SV-min also exhibit certain underestimation in the heavy mass region.
Though the Skyrme HFB results with UNEDF1 improve the description in $A\gtrsim 170$, they apparently underestimate the masses of a few light and medium-mass nuclei, and show significant overestimation at $A\approx 155$.
The comparisons in Fig.~\ref{fig3} demonstrate that the descriptions from DRHBc with PC-PK1 and PC-L3R are qualitatively superior.

\begin{figure}[htbp]
\centering
\includegraphics[width=3.0in]{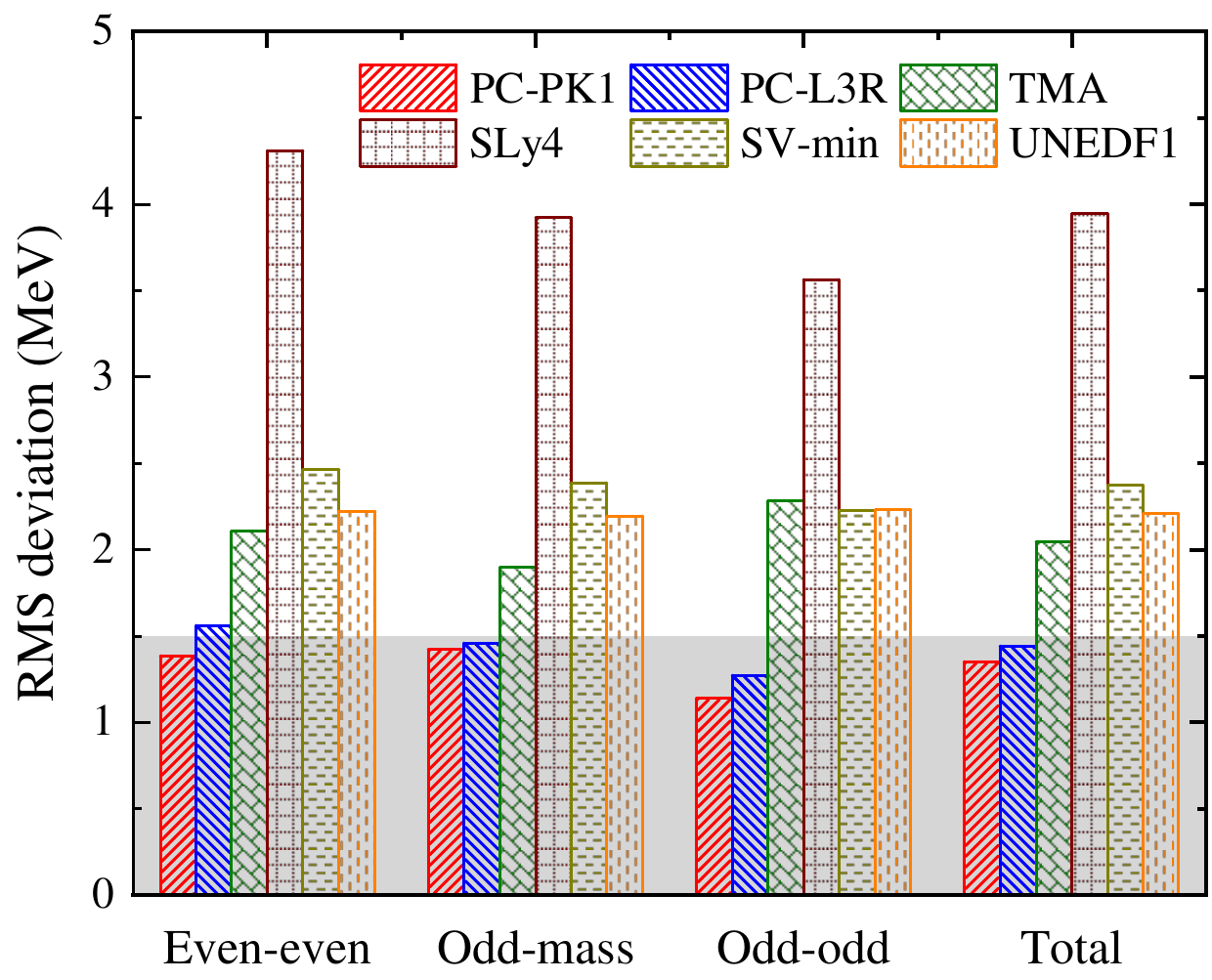}
\caption{The RMS deviations between the new mass data listed in Table \ref{mass-table} and different theoretical results. The RMS deviations for even-even, odd-mass and odd-odd nuclei are presented separately.}
\label{fig4}
\end{figure}

To make a further comparison, we show in Fig.~\ref{fig4} the RMS deviations between the 296 new mass data and the above theoretical results.
The RMS deviations for even-even, odd-mass and odd-odd nuclei are also computed and presented separately in Fig.~\ref{fig4}.
It can be found that both the DRHBc descriptions with PC-PK1 and PC-L3R can achieve accuracies better than 1.5 MeV for all data sets, with one exception being DRHBc+PC-L3R for even-even nuclei.
In contrast, the accuracies in other four density-functional descriptions are in general worse than 2 MeV.
Overall, the odd-even effects in the accuracy are not very significant in the DRHBc results, with even slightly better description for odd-odd nuclei.
This is, however, not the case for the RMF+BCS description, which deteriorates obviously for odd-odd nuclei.
Furthermore, the DRHBc descriptions with PC-PK1 and PC-L3R for odd nuclei are expected to be improved by strictly incorporating nuclear magnetism \cite{Pan2024PLB}.
Instead of self-consistent calculations, the Skyrme HFB results for odd nuclei are obtained from interpolations using the masses and average pairing gaps of neighboring even-even nuclei \cite{Massexp}.
As expected, the Skyrme HFB descriptions with SV-min and UNEDF1 show marginal odd-even differences.
In contrast, it seems strange that from even-even to odd-mass and then to odd-odd nuclei, the SLy4 description gradually improves.
It should also be noted that the number of mass data here is not large enough to confirm whether the odd-even features observed in these theoretical results are common across the nuclear chart.
Finally, the accuracies in describing the 296 new masses--1.35, 2.04, 3.95, 2.37, and 2.21 MeV for PC-PK1, TMA, SLy4, SV-min, and UNEDF1, respectively--are found to be generally consistent with those obtained for all available masses of even-$Z$ nuclei: 1.43 MeV for PC-PK1, 2.06 MeV for TMA, 5.28 MeV for SLy4, 3.39 MeV for SV-min, and 1.93 MeV for UNEDF1 \cite{Guo2024ADNDT}.
Moreover, even within the spherical RHB framework, PC-PK1 and PC-L3R are the only two relativistic density functionals that reproduce experimental masses with rms deviations below 8 MeV \cite{Xia2018ADNDT,Liu2024ADNDT}.
Given the superiority of PC-PK1 and PC-L3R, the complete DRHBc mass table including both even-$Z$ and odd-$Z$ nuclei in the near future is desirable, and further large-scale DRHBc+PC-L3R calculations are worth pursuing.

\begin{figure*}[!htb]
\centering
\includegraphics[width=5in]{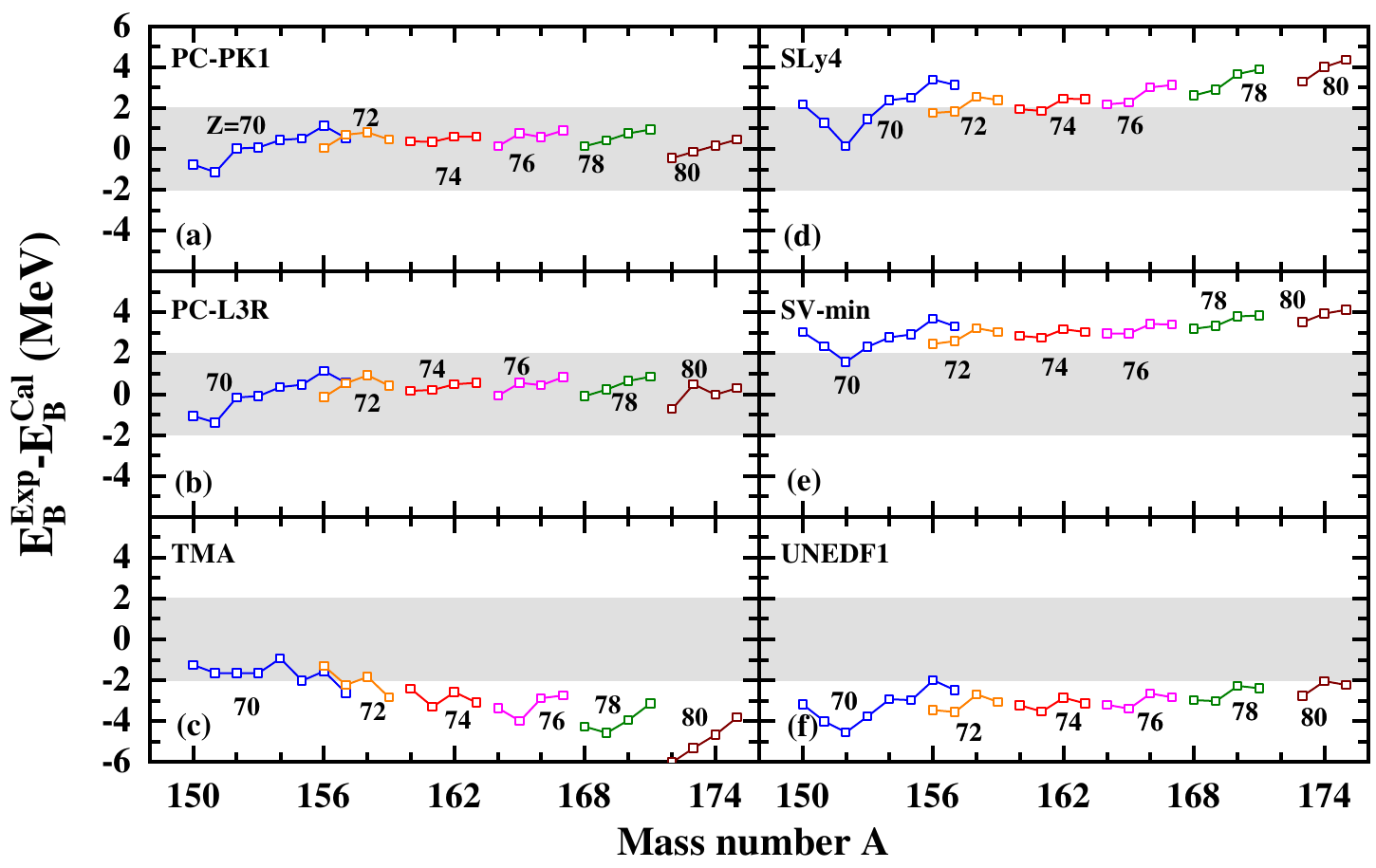}
\caption{Same as Fig. \ref{fig3} but for the isotopic chains with $Z=70$, 72, 74, 76, 78, and 80.}
\label{fig5}
\end{figure*}

One can see from Fig.~\ref{fig3} that the superiority of PC-PK1 and PC-L3R mainly comes from the better description for the nuclei with $A>150$ compared to other density functionals.
Therefore, a detailed comparison for the description of the isospin dependence of nuclear masses in this region would be illuminating.
In Fig.~\ref{fig5}, the mass differences between theoretical and experimental values are presented for even-$Z$ nuclei with $70\le Z\le 80$.
In case masses of certain nuclei located in the middle of an isotopic chain are absent from the dataset of new measurements, we resort to AME2020 for completeness.
It can be found in Fig.~\ref{fig5} that only the accuracy of the DRHBc description in this region is always better than 2 MeV, while other density-functional descriptions show systematic deviations from the data.
Furthermore, the DRHBc description along these isotopic chains is almost steady, with slight, approximately linear isospin dependence, which is in contrast to many obvious staggering behaviors by other descriptions.
Another feature observed in Fig.~\ref{fig5} is that the nucleus with $N=82$ and $Z=70$ is basically described as over bound compared to its neighboring isotopes.
This is a well-known weakness of both non-relativistic and relativistic density functional theories in describing magic nuclei \cite{Zhang2022ADNDT}.
It can be concluded from the above discussions that the DRHBc theory provides not only an overall high accuracy but also a robust description of isospin dependence for nuclear masses, and the results from PC-PK1 and PC-L3R are qualitatively similar.

\begin{figure}[htbp]
\centering	
\includegraphics[width=8cm]{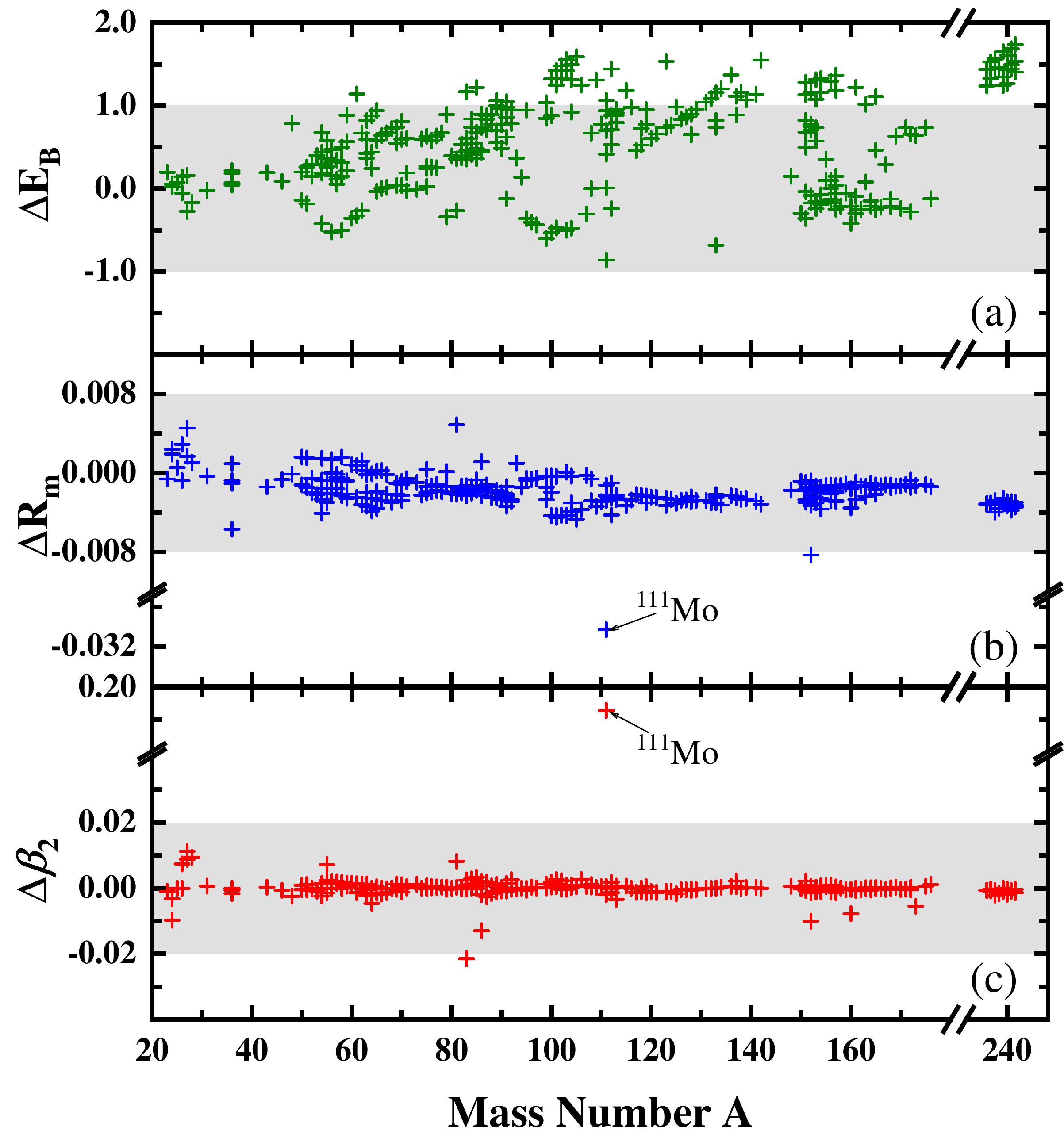}
\caption{Differences between the DRHBc results with PC-PK1 and PC-L3R for binding energies $\Delta E_B = E_B^{\text{PC-PK1}} - E_B^{\text{PC-L3R}}$ (a), RMS matter radii $\Delta R_{m} = R_{m}^{\text{PC-PK1}} - R_{m}^{\text{PC-L3R}}$ (b) and quadrupole deformations $\Delta \beta_2 = \beta_2^{\text{PC-PK1}} - \beta_2^{\text{PC-L3R}}$ (c) as functions of mass number $A$.}
\label{fig6}
\end{figure}

To make a quantitative comparison, the differences between the DRHBc results with PC-PK1 and PC-L3R for binding energies $\Delta E_B = E_B^{\text{PC-PK1}} - E_B^{\text{PC-L3R}}$, rms matter radii $\Delta R_{m} = R_{m}^{\text{PC-PK1}} - R_{m}^{\text{PC-L3R}}$, and quadrupole deformations $\Delta \beta_2 = \beta_2^{\text{PC-PK1}} - \beta_2^{\text{PC-L3R}}$ are shown in Fig.~\ref{fig6}.
For the binding energies in Fig.~\ref{fig6}(a), among these 296 nuclei, the $\Delta E_B$ of 232 nuclei locate within $-1.0 < \Delta E_B < 1.0$ MeV, while the $\Delta E_B$ of 64 nuclei locate within $1.0 < \Delta E_B < 2.0$ MeV.
Most values of $\Delta E_B$ are positive, indicating that PC-PK1 generally describes these nuclei as more bound than PC-L3R. 
The rms matter radii shown in Fig.~\ref{fig6}(b) reveal a clear trend of decreasing $\Delta R_m$ with increasing mass number $A$, with only a few exceptions.
Notably, the majority of $\Delta R_m$ values are negative, consistent with the general expectation that more strongly bound systems exhibit more compact density distributions.
While for most nuclei $\Delta R_m$ values are confined within $\pm 0.008$ fm, two outliers emerge: one barely beyond $-0.008$ fm for $^{152}$Pr and the other reaching $-0.031$ fm for $^{111}$Mo.
For the quadrupole deformation in Fig.~\ref{fig6}(c), the $\Delta \beta_2$ in 291 nuclei are within 0.01, and almost all other nuclei show slightly larger values within 0.02, except for $^{83}$Zr and $^{111}$Mo.
The $\Delta \beta_2$ for $^{83}$Zr is only $-0.021$, but that for $^{111}$Mo reaches $0.193$, which corresponds to the large $|\Delta R_{m}|$ shown in Fig.~\ref{fig6}(b).

\begin{figure}[htbp]
\centering	
\includegraphics[width=8cm]{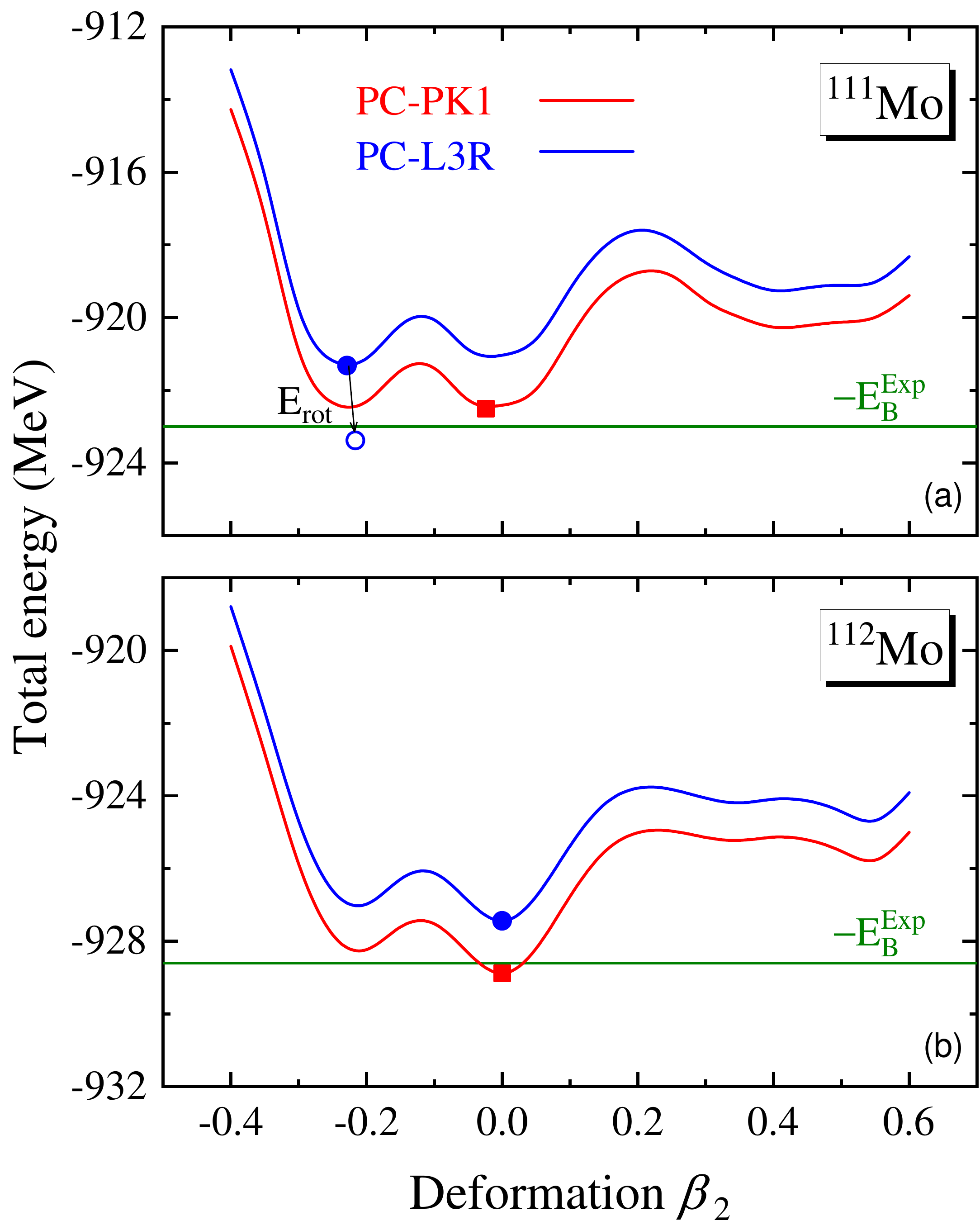}
\caption{Potential energy curves (PECs) of $^{111}$Mo (a) and $^{112}$Mo (b) in constrained DRHBc calculations with PC-PK1 (red line) and PC-L3R (blue line).
The ground state is shown with the filled square (circle) from the calculations with PC-PK1 (PC-L3R).
For the case with $|\beta_2|>0.05$, the ground state after including the rotational correction energy $E_{\rm tot}$ is shown with the open symbol.
The green line represents the available data of $-E^{\rm Exp}_{B}$.
}
\label{fig7}
\end{figure}

To understand the large deviations between the PC-PK1 and PC-L3R results for $^{111}$Mo, in Fig.~\ref{fig7}, the potential energy curves (PECs) from $^{111}$Mo in constrained calculations with PC-PK1 and PC-L3R are shown.
For comparison, the corresponding results of its neighboring isotope $^{112}$Mo are also shown.
The ground state is shown with the filled square (circle) from the calculations with PC-PK1 (PC-L3R).
As has been demonstrated in Refs.~\cite{Zhao2010Phys.Rev.C54319,ZhangKY2020PRC102,Pan2022PRC}, for PC-PK1, the rotational correction plays an important role in improving its mass description of deformed nuclei.
Therefore, the ground state after including the rotational correction energy $E_{\text{rot}}$ is also presented with the open symbol.
Considering that the cranking approximation adopted to calculate $E_{\text{rot}}$ in Eq.~\eqref{erot} is not suitable for spherical and weakly deformed nuclei, we only calculate $E_{\text{rot}}$ when $|\beta_2|>0.05$, and take it as zero when $|\beta_2|<0.05$.
A more proper treatment for the correction energies in the nuclei with $|\beta_2|<0.05$ can be achieved by using the collective Hamiltonian method \cite{SunW2022CPC} in future work.

For $^{111}$Mo in Fig.~\ref{fig7}(a), in both the PECs from the calculations with PC-PK1 and PC-L3R, there are three minima, i.e., one near-spherical minimum and two well-deformed minima.
In each curve, the difference between the oblate and near-spherical minima is within 0.25 MeV, while the excitation energy of the prolate minimum is about 2 MeV.
The ground state from the PC-PK1 calculations is the near-spherical minimum with $\beta_2=-0.024$, while the ground state from the PC-L3R calculations is the oblate minimum with $\beta_2=-0.216$.
This corresponds to the large $\Delta\beta_2$ in Fig.~\ref{fig6}(c).
In addition, for the near-spherical ground state from PC-PK1, the rotational correction energy is taken as zero.
For the well-deformed ground state from PC-L3R, the rotational correction energy is $E_{\rm rot}=-2.06$ MeV, with which the result becomes more deeply bound than that from PC-PK1.
Finally, the binding energies from the PC-PK1 and PC-L3R calculations are similar, both well reproducing the experimental value for $^{111}$Mo.

For $^{112}$Mo, the shape of the PEC is similar to that for $^{111}$Mo, still exhibiting three minima.
However, here the ground states from both PC-PK1 and PC-L3R are the spherical minima, while the excitation energies of the oblate and prolate minima are about 0.5 and 3 MeV, respectively.
In both results from PC-PK1 and PC-L3R, the significant differences in deformations and the small differences in total energies indicate the shape coexistence in $^{111,112}$Mo.

\section{Summary}\label{section4}

In this work, the newly measured masses for 296 nuclides from 40 references published between 2021 and 2024, subsequent to the release of the latest Atomic Mass Evaluation, are compiled.
While most of the new data are consistent with AME2020, for 73 nuclides the deviations exceed the uncertainties.
The new masses are calculated using the DRHBc theory with the PC-PK1 and PC-L3R density functionals, and compared with results from RMF+BCS calculations with TMA, and Skyrme HFB calculations with SLy4, SV-min, and UNEDF1.
The DRHBc calculations with both PC-PK1 and PC-L3R reproduce fairly well the data with a RMS deviation below 1.5 MeV, demonstrating a clear advantage over other models in mass predictions.
Taking the even-$Z$ nuclei with $70 \le Z \le 80$ as examples, the DRHBc calculations provide not only an overall high accuracy but also a robust description of isospin dependence for nuclear masses.
A quantitative comparison between PC-PK1 and PC-L3R results for the 296 nuclides shows that their differences in binding energies are generally below 1.0 MeV, those in RMS matter radii within 0.008 fm, and those in quadrupole deformations within 0.01.
The largest discrepancies are found in $^{111}$Mo, attributed to the competing oblate and near-spherical minima with similar energies in the potential energy curve.
The significant differences in deformation and the small differences in energies between the two minima indicate possible shape coexistence in $^{111}$Mo and $^{112}$Mo.

These results strengthen the confidence in the DRHBc predictions of nuclear masses.
Continued progress would benefit from additional experimental mass measurements and the establishment of the complete DRHBc mass table with PC-PK1.
Moreover, large-scale DRHBc calculations with PC-L3R are also highly promising.
To further improve the mass description, the inclusion of triaxial deformation, the rigorous treatment of nuclear magnetism, and the beyond-mean-field extensions such as collective Hamiltonian approaches, are being pursued within the DRHBc framework.

\bibliographystyle{unsrt}
\bibliography{Ref1}{}

\begin{appendix}

\setcounter{table}{0}
\renewcommand{\thetable}{A\arabic{table}}
\section{}\label{append}

\setlength{\tabcolsep}{-1pt}
\begin{longtable*}{w{c}{1cm} w{r}{4.5cm} w{r}{4cm} w{r}{2.5cm}}
\caption{Newly measured masses of 296 nuclides from 2021 to 2024, in comparison with the AME2020 data \cite{WangM2021CPC}, where the values in brackets represent uncertainties and \# denotes extrapolated values. Inconsistent data after considering uncertainties are highlighted in red and bold. The unit of the data is MeV.}\label{mass-table}\\\hline\hline
Nucl. & \multicolumn{1}{c}{New} & \multicolumn{1}{c}{AME2020~\cite{WangM2021CPC}}   & $E_{\text{B}}^{\text{New}} - E_{\text{B}}^{\text{AME}}$   \\\hline
\endfirsthead
{{\bfseries \tablename\ \thetable{} -- continued}} \\\hline
Nucl. &  \multicolumn{1}{c}{New}   &  \multicolumn{1}{c}{AME2020~\cite{WangM2021CPC}}  &  \multicolumn{1}{c}{$E_{\text{B}}^{\text{New}} - E_{\text{B}}^{\text{AME}}$}   \\\hline
\endhead
\endfoot   \hline \hline
\endlastfoot
$^{24}$Ne   &~191.84268~(0.06900)     \cite{jacobs2023improved}                  &~191.84028~(0.00051)                      &~   0.00288         \\ \hline
$^{25}$Ne   &~195.98389~(0.08000)     \cite{jacobs2023improved}                  &~195.99599~(0.02905)                      &~  $-$0.01110         \\ \hline
$^{26}$Ne   &~201.57490~(0.08000)     \cite{jacobs2023improved}                  &~201.55030~(0.01843)                      &~   0.02521         \\ \hline
$^{23}$Si   &~151.15127~(0.11900)     \cite{YUY2024PRL133}                		 &~$^{\#}$150.74200~(0.50000)               &~   0.40927         \\ \hline
$^{24}$Si   &~172.01359~(0.03700)     \cite{Puentes106PRC}                       &~172.00580~(0.01947)                      &~  $-$0.00781         \\ \hline
$^{26}$P    &~187.12192~(0.01100)     \cite{YUY2024PRL133}                       &~$^{\#}$187.14800~(0.50000)               &~  $-$0.02608         \\ \hline
$^{27}$P    &~206.86183~(0.00600)     \cite{Yandow2023PRC}                       &~206.85023~(0.00900)                      &~   0.01252         \\ \hline
$^{27}$S    &~187.99095~(0.03900)     \cite{YUY2024PRL133}                       &~$^{\#}$187.92000~(0.40000)               &~   0.07095         \\ \hline
$^{28}$S    &{\bf\color{red}{~209.11792~(0.01400)}} \cite{YUY2024PRL133}         &{\bf\color{red}{~209.40615~(0.16000)}}    &~  $-$0.28823         \\ \hline
$^{36}$S    &~308.71486~(0.01900)     \cite{Surbrook2021PRC}                     &~308.71404~(0.00019)                      &~   0.00082         \\ \hline
$^{36}$Cl   &~306.79048~(0.00400)     \cite{Surbrook2021PRC}                     &~306.78956~(0.00004)                      &~   0.00092         \\ \hline
$^{31}$Ar   &~224.83964~(0.01600)     \cite{YUY2024PRL133}                       &~$^{\#}$224.81200~(0.20000)               &~   0.02764         \\ \hline
$^{36}$Ar   &~306.71773~(0.02700)     \cite{Surbrook2021PRC}                     &~306.71675~(0.00003)                      &~   0.00098         \\ \hline
$^{36}$K    &~293.12102~(0.00300)     \cite{Surbrook2021PRC}                     &~293.12004~(0.00033)                      &~   0.00098         \\ \hline
$^{36}$Ca   &~281.40524~(0.05600)     \cite{Surbrook2021PRC}                     &~281.37168~(0.04000)                      &~   0.03356         \\ \hline
$^{54}$Ca   &~445.32420~(0.01200)     \cite{Porter2022PRC106}                    &~445.36482~(0.04844)                      &~  $-$0.04060         \\ \hline
$^{50}$Sc   &~431.67480~(0.00250)     \cite{Lebit2021PRL}                        &~431.67374~(0.00252)                      &~   0.00106         \\ \hline
$^{51}$Sc   &~438.45941~(0.00250)     \cite{Lebit2021PRL}                        &~438.45829~(0.00252)                      &~   0.00112         \\ \hline
$^{52}$Sc   &~443.80392~(0.00300)     \cite{Lebit2021PRL}                        &~443.80281~(0.00307)                      &~   0.00111         \\ \hline
$^{53}$Sc   &~450.12063~(0.01700)     \cite{Lebit2021PRL}                        &~450.12012~(0.01770)                      &~   0.00051         \\ \hline
$^{54}$Sc   &~453.86094~(0.01800)     \cite{Lebit2021PRL}                        &~453.85982~(0.01397)                      &~   0.00112         \\ \hline
$^{55}$Sc   &~458.33625~(0.06200)     \cite{Lebit2021PRL}                        &~458.33532~(0.06241)                      &~   0.00093         \\ \hline
$^{43}$Ti   &{\bf\color{red}{~359.15846~(0.00420)}}    \cite{WangM2022PRC}       &{\bf\color{red}{~359.17063~(0.00572)}}    &~  $-$0.01217         \\ \hline
$^{52}$Ti   &~451.97665~(0.00220)    \cite{Porter2022PRC106}                     &~451.97460~(0.00275)                      &~   0.00205         \\ \hline
$^{54}$Ti   &~464.37867~(0.02700)    \cite{Porter2022PRC106}                     &~464.38335~(0.01584)                      &~  $-$0.00468         \\ \hline
$^{55}$Ti   &~468.53898~(0.00570)    \cite{Porter2022PRC106}                     &~468.54333~(0.02888)                      &~  $-$0.00435         \\ \hline
$^{56}$Ti   &~474.19149~(0.00740)    \cite{Iimura2023PRL}                        &~474.20517~(0.10020)                      &~  $-$0.01368         \\ \hline
$^{54}$V    &~467.75739~(0.01000)    \cite{Porter2022PRC106}                     &~467.75546~(0.01118)                      &~   0.00193         \\ \hline
$^{55}$V    &~475.06790~(0.00660)    \cite{Porter2022PRC106}                     &~475.05365~(0.02701)                      &~   0.01425         \\ \hline
$^{56}$V    &~480.26061~(0.00620)    \cite{Iimura2023PRL}                        &~480.18323~(0.17588)                      &~   0.07738         \\ \hline
$^{57}$V    &~486.45532~(0.01500)    \cite{Iimura2023PRL}                        &~486.50621~(0.08477)                      &~  $-$0.05089         \\ \hline
$^{58}$V    &{\bf\color{red}{~490.44973~(0.00560)}}    \cite{Iimura2023PRL}      &{\bf\color{red}{~490.57305~(0.09582)}}    &~  $-$0.12332         \\ \hline
$^{59}$V    &{\bf\color{red}{~496.01714~(0.00280) }}   \cite{Iimura2023PRL}      &{\bf\color{red}{~495.82440~(0.13740)}}    &~   0.19274         \\ \hline
$^{46}$Cr   &~381.98284~(0.00260)    \cite{WangM2022PRC}                         &~381.97587~(0.01145)                      &~   0.00697         \\ \hline
$^{56}$Cr   &{\bf\color{red}{~488.51374~(0.00530)}}    \cite{Iimura2023PRL}      &{\bf\color{red}{~488.50261~(0.00058)}}    &~   0.01113         \\ \hline
$^{57}$Cr   &~493.82605~(0.01200)    \cite{Iimura2023PRL}                        &~493.81379~(0.00186)                      &~   0.01226         \\ \hline
$^{58}$Cr   &{\bf\color{red}{~501.36086~(0.00370)}}    \cite{Iimura2023PRL}      &{\bf\color{red}{~501.35192~(0.00298)}}    &~   0.00894         \\ \hline
$^{59}$Cr   &~505.54567~(0.00700)    \cite{Silwal2022PLB}                        &~505.54737~(0.00067)                      &~  $-$0.00170         \\ \hline
$^{61}$Cr   &~516.07229~(0.01400)    \cite{Silwal2022PLB}                        &~516.07058~(0.00186)                      &~   0.00171         \\ \hline
$^{62}$Cr   &~522.50160~(0.01900)    \cite{Silwal2022PLB}                        &~522.49800~(0.00345)                      &~   0.00360         \\ \hline
$^{63}$Cr   &~525.92191~(0.01800)    \cite{Silwal2022PLB}                        &~525.89501~(0.07266)                      &~   0.02690         \\ \hline
$^{64}$Cr   &~531.36222~(0.02600)    \cite{Silwal2022PLB}                        &~531.42801~(0.29994)                      &~  $-$0.06579         \\ \hline
$^{65}$Cr   &~534.06853~(0.04500)    \cite{Silwal2022PLB}                        &~$^{\#}$534.17000~(0.20000)               &~  $-$0.10147         \\ \hline
$^{48}$Mn   &~397.15638~(0.00290)    \cite{WangM2022PRC}                         &~397.16124~(0.00670)                      &~  $-$0.00486         \\ \hline
$^{57}$Mn   &~498.00377~(0.01500)    \cite{Iimura2023PRL}                        &~497.99273~(0.00151)                      &~   0.01104         \\ \hline
$^{58}$Mn   &~504.40808~(0.00590)    \cite{Iimura2023PRL}                        &~504.40534~(0.00270)                      &~   0.00274         \\ \hline
$^{59}$Mn   &~512.17509~(0.00500)    \cite{Iimura2023PRL}                        &~512.17442~(0.00233)                      &~   0.00067         \\ \hline
$^{50}$Fe   &~417.70212~(0.00290)    \cite{WangM2022PRC}                         &~417.70134~(0.00838)                      &~   0.00078         \\ \hline
$^{51}$Fe   &~431.49953~(0.01590)    \cite{WangM2022PRC}                         &~431.48538~(0.00140)                      &~   0.01415         \\ \hline
$^{63}$Fe   &~543.78875~(0.00540)    \cite{Porter2022PRC105}                     &~543.78764~(0.00430)                      &~   0.00111         \\ \hline
$^{64}$Fe   &~551.19466~(0.00530)    \cite{Porter2022PRC105}                     &~551.19288~(0.00502)                      &~   0.00178         \\ \hline
$^{65}$Fe   &~555.51467~(0.00840)    \cite{Porter2022PRC105}                     &~555.51256~(0.00511)                      &~   0.00212         \\ \hline
$^{66}$Fe   &~562.42828~(0.01000)    \cite{Porter2022PRC105}                     &~562.43382~(0.00410)                      &~  $-$0.00554         \\ \hline
$^{67}$Fe   &{\bf\color{red}{~566.45339~(0.00870)}}    \cite{Porter2022PRC105}   &{\bf\color{red}{~566.14571~(0.00382)}}    &~   0.30768         \\ \hline
$^{68}$Fe   &~572.61080~(0.00560)    \cite{Porter2022PRC105}                     &~$^{\#}$572.42400~(0.00300)               &~   0.18680         \\ \hline
$^{69}$Fe   &~576.08521~(0.01100)    \cite{Porter2022PRC105}                     &~$^{\#}$575.80500~(0.00300)               &~   0.28021         \\ \hline
$^{70}$Fe   &~581.70552~(0.01200)    \cite{Porter2022PRC105}                     &~$^{\#}$581.56000~(0.00400)               &~   0.14552         \\ \hline
$^{69}$Co   &~586.18394~(0.08600)    \cite{Canete2021PRC}                        &~586.18303~(0.08570)                      &~   0.00091         \\ \hline
$^{70}$Co   &~590.39525~(0.01100)    \cite{Canete2021PRC}                        &~590.39386~(0.01099)                      &~   0.00139         \\ \hline
$^{54}$Ni   &{\bf\color{red}{~453.23241~(0.00270)}}     \cite{WangM2022PRC}      &{\bf\color{red}{~453.22377~(0.00466)}}    &~   0.00864         \\ \hline
$^{74}$Ni   &~623.82461~(0.00350)     \cite{giraud2022mass}                      &~$^{\#}$624.07321~(0.20000)               &~  $-$0.21739         \\ \hline
$^{75}$Ni   &~627.50042~(0.01470)     \cite{giraud2022mass}                      &~$^{\#}$627.68452~(0.20000)               &~  $-$0.17458         \\ \hline
$^{56}$Cu   &~467.92996~(0.00600)     \cite{WangM2022PRC}                        &~467.93548~(0.00011)                      &~  $-$0.00552         \\ \hline
$^{76}$Cu   &{\bf\color{red}{~641.74496~(0.00200)}} \cite{giraud2022mass}        &{\bf\color{red}{~641.71516~(0.00091)}}    &~   0.03122         \\ \hline
$^{77}$Cu   &~647.66637~(0.00470)     \cite{giraud2022mass}                      &~647.66767~(0.00121)                      &~   0.00011         \\ \hline
$^{78}$Cu   &~651.66188~(0.00750)     \cite{giraud2022mass}                      &~651.66518~(0.01333)                      &~  $-$0.00234         \\ \hline
$^{58}$Zn   &~486.91570~(0.03600)     \cite{wang2023mass}                        &~486.96770~(0.05000)                      &~  $-$0.04921         \\ \hline
$^{79}$Zn   &~667.59701~(0.00310)     \cite{giraud2022mass}                      &~667.59751~(0.00223)                      &~   0.00099         \\ \hline
$^{60}$Ga   &~500.06204~(0.04600)     \cite{PhysRevC.104.065803}                 &~$^{\#}$499.61804~(0.20000)               &~   0.44204         \\ \hline
$^{61}$Ga   &~515.26735~(0.02100)     \cite{PhysRevC.104.065803}                 &~515.22935~(0.03799)                      &~   0.03504         \\ \hline
$^{62}$Ga   &~528.16266~(0.01400)     \cite{PhysRevC.104.065803}                 &~528.16266~(0.00064)                      &~   0.00668         \\ \hline
$^{63}$Ga   &{\bf\color{red}{~540.80497~(0.01400)}}  \cite{PhysRevC.104.065803}  &{\bf\color{red}{~540.78738~(0.00130)}}    &~   0.01759         \\ \hline
$^{83}$Ga   &{\bf\color{red}{~695.02193~(0.07100)}}  \cite{XianW2024PRC}         &{\bf\color{red}{~694.92377~(0.00261)}}    &~   0.09816         \\ \hline
$^{84}$Ga   &~723.94076~(0.03300)     \cite{XianW2024PRC}                        &~697.83210~(0.02981)                      &~   0.08996         \\ \hline
$^{62}$Ge   &~517.67738~(0.04600)     \cite{wang2023mass}                        &~$^{\#}$517.52839~(0.14000)               &~   0.16339         \\ \hline
$^{63}$Ge   &{\bf\color{red}{~530.43770~(0.01500)}}\cite{zhou2023Mass}           &{\bf\color{red}{~530.38070~(0.03726)}}  &~   0.05855         \\ \hline
$^{82}$Ge   &{\bf\color{red}{~702.32013~(0.01600)}}     \cite{XianW2024PRC}      &{\bf\color{red}{~702.22805~(0.00224)}}    &~   0.09208         \\ \hline
$^{83}$Ge   &~705.95446~(0.09200)     \cite{XianW2024PRC}                        &~705.86073~(0.00243)                      &~   0.09373         \\ \hline
$^{84}$Ge   &{\bf\color{red}{~711.19929~(0.03000)}}     \cite{XianW2024PRC}      &{\bf\color{red}{~711.10405~(0.00317)}}    &~   0.09524         \\ \hline
$^{85}$Ge   &{\bf\color{red}{~714.23992~(0.05500)}}     \cite{XianW2024PRC}      &{\bf\color{red}{~714.15036~(0.00373)}}    &~   0.08956         \\ \hline
$^{86}$Ge   &~718.79285~(0.01700)     \cite{XianW2024PRC}                        &~718.49818~(0.43780)                      &~   0.29467         \\ \hline
$^{64}$As   &~530.45873~(0.11000)     \cite{wang2023mass}                        &~$^{\#}$530.27873~(0.20300)               &~   0.15473         \\ \hline
$^{65}$As   &~{\bf\color{red}{~545.62604~(0.04200)}}\cite{wang2023mass}          &{\bf\color{red}{~545.75704~(0.08477)}}    &~  $-$0.12924         \\ \hline
$^{69}$As   &~594.24208~(0.02900)     \cite{xing2023isochronous}                 &~594.21568~(0.03200)                      &~   0.02641         \\ \hline
$^{75}$As   &~652.56912~(0.04300)     \cite{Horana2022PRC}                       &~652.56561~(0.03200)                      &~   0.00351         \\ \hline
$^{76}$As   &~659.89714~(0.07500)     \cite{XianW2024PRC}                        &~659.89411~(0.00088)                      &~   0.00303         \\ \hline
$^{82}$As   &{\bf\color{red}{~706.23135~(0.03300)}}     \cite{XianW2024PRC}      &{\bf\color{red}{~706.13605~(0.00373)}}    &~   0.09530         \\ \hline
$^{83}$As   &{\bf\color{red}{~713.86658~(0.01100)}}     \cite{XianW2024PRC}      &{\bf\color{red}{~713.77128~(0.00373)}}    &~   0.09530         \\ \hline
$^{84}$As   &{\bf\color{red}{~718.12251~(0.04100)}}     \cite{XianW2024PRC}      &{\bf\color{red}{~718.02683~(0.00317)}}    &~   0.09568         \\ \hline
$^{85}$As   &{\bf\color{red}{~723.53784~(0.02800)}}    \cite{XianW2024PRC}       &{\bf\color{red}{~723.43373~(0.00308)}}    &~   0.10411         \\ \hline
$^{86}$As   &{\bf\color{red}{~727.37847~(0.02000)}}     \cite{XianW2024PRC}      &{\bf\color{red}{~727.27805~(0.00345)}}    &~   0.10042         \\ \hline
$^{87}$As   &{\bf\color{red}{~732.10410~(0.02900)}}     \cite{XianW2024PRC}      &~732.00513~(0.00299)                      &~   0.09897         \\ \hline
$^{88}$As   &~735.23653~(0.03100)     \cite{XianW2024PRC}                        &~$^{\#}$734.88800~(0.00200)               &~   0.34853         \\ \hline
$^{89}$As   &~739.31766~(0.01700)     \cite{XianW2024PRC}                        &~$^{\#}$739.05600~(0.00300)               &~   0.26166         \\ \hline
$^{66}$Se   &~548.09108~(0.06100)     \cite{wang2023mass}                        &~$^{\#}$547.76908~(0.20000)               &~   0.29108         \\ \hline
$^{67}$Se   &~560.72939~(0.02000)     \cite{zhou2023Mass}                        &~560.75880~(0.06707)                      &~  $-$0.02942         \\ \hline
$^{70}$Se   &~600.32032~(0.00260)     \cite{PhysRevC.103.034319}                 &~600.32237~(0.00158)                      &~  $-$0.00205         \\ \hline
$^{71}$Se   &~609.61263~(0.02300)     \cite{PhysRevC.103.034319}                 &~609.61030~(0.00279)                      &~   0.00232         \\ \hline
$^{77}$Se   &~669.49293~(0.05800)     \cite{XianW2024PRC}                        &~669.49119~(0.00006)                      &~   0.00174         \\ \hline
$^{82}$Se   &{\bf\color{red}{~712.93168~(0.03400)}}     \cite{XianW2024PRC}      &{\bf\color{red}{~712.84218~(0.00047)}}    &~   0.08950         \\ \hline
$^{84}$Se   &{\bf\color{red}{~727.43924~(0.04700)}}     \cite{XianW2024PRC}      &{\bf\color{red}{~727.33865~(0.00196)}}    &~   0.10059         \\ \hline
$^{85}$Se   &{\bf\color{red}{~731.97507~(0.02000)}}     \cite{XianW2024PRC}      &{\bf\color{red}{~731.87588~(0.00261)}}    &~   0.09919         \\ \hline
$^{86}$Se   &{\bf\color{red}{~738.13510~(0.01200)}}     \cite{XianW2024PRC}      &{\bf\color{red}{~738.03673~(0.00252)}}    &~   0.09837         \\ \hline
$^{87}$Se   &{\bf\color{red}{~742.12933~(0.03200)}}     \cite{XianW2024PRC}      &{\bf\color{red}{~742.03100~(0.00224)}}    &~   0.09833         \\ \hline
$^{88}$Se   &{\bf\color{red}{~747.65896~(0.02100)}}     \cite{XianW2024PRC}      &{\bf\color{red}{~747.56040~(0.00336)}}    &~   0.09856         \\ \hline
$^{89}$Se   &{\bf\color{red}{~750.83829~(0.02100)}}     \cite{XianW2024PRC}      &{\bf\color{red}{~750.73991~(0.00373)}}    &~   0.09838         \\ \hline
$^{90}$Se   &~755.80262~(0.04300)     \cite{XianW2024PRC}                        &~755.61905~(0.32975)                      &~   0.18357         \\ \hline
$^{91}$Se   &~758.26115~(0.02600)     \cite{XianW2024PRC}                        &~758.47028~(0.43315)                      &~  $-$0.20913         \\ \hline
$^{71}$Br   &~602.16435~(0.01600)     \cite{PhysRevC.103.034319}                 &~602.18387~(0.00540)                      &~  $-$0.01952         \\ \hline
$^{73}$Br   &~625.44887~(0.02700)     \cite{xing2023isochronous}                 &~625.46986~(0.00674)                      &~  $-$0.02099         \\ \hline
$^{85}$Br   &{\bf\color{red}{~737.37589~(0.01200)}}     \cite{XianW2024PRC}      &{\bf\color{red}{~737.25537~(0.00308)}}    &~   0.12052         \\ \hline
$^{86}$Br   &{\bf\color{red}{~742.48772~(0.01700)}}     \cite{XianW2024PRC}      &{\bf\color{red}{~742.38347~(0.00308)}}    &~   0.10425         \\ \hline
$^{89}$Br   &{\bf\color{red}{~759.34161~(0.04800)}}     \cite{XianW2024PRC}      &{\bf\color{red}{~759.23944~(0.00326)}}    &~   0.10217         \\ \hline
$^{90}$Br   &{\bf\color{red}{~763.13784~(0.04300)}}     \cite{XianW2024PRC}      &{\bf\color{red}{~763.03679~(0.00336)}}    &~   0.10105         \\ \hline
$^{91}$Br   &{\bf\color{red}{~768.31827~(0.02500)}}     \cite{XianW2024PRC}      &{\bf\color{red}{~768.21510~(0.00354)}}    &~   0.10317         \\ \hline
$^{92}$Br   &{\bf\color{red}{~771.52440~(0.06400)}}     \cite{XianW2024PRC}      &{\bf\color{red}{~771.41193~(0.00671)}}    &~   0.11247         \\ \hline
$^{70}$Kr   &~578.14976~(0.14000)    \cite{wang2023mass}                         &~$^{\#}$577.92976~(0.20000)               &~   0.22976         \\ \hline
$^{71}$Kr   &{\bf\color{red}{~590.95707~(0.02400)}}\cite{wang2023mass}           &{\bf\color{red}{~591.22807~(0.12877)}}    &~  $-$0.26923         \\ \hline
$^{75}$Kr   &~641.51091~(0.02600)     \cite{xing2023isochronous}                 &~641.50991~(0.00810)                      &~   0.00292         \\ \hline
$^{89}$Kr   &{\bf\color{red}{~766.82133~(0.01200)}}     \cite{XianW2024PRC}      &{\bf\color{red}{~766.71861~(0.00214)}}    &~   0.10272         \\ \hline
$^{91}$Kr   &{\bf\color{red}{~777.40029~(0.05700)}}     \cite{XianW2024PRC}      &{\bf\color{red}{~777.29942~(0.00224)}}    &~   0.10087         \\ \hline
$^{92}$Kr   &{\bf\color{red}{~783.27362~(0.08500)}}     \cite{XianW2024PRC}      &{\bf\color{red}{~783.16610~(0.00270)}}    &~   0.10752         \\ \hline
$^{90}$Rb   &{\bf\color{red}{~776.92679~(0.05300)}}     \cite{XianW2024PRC}      &{\bf\color{red}{~776.83736~(0.00645)}}    &~   0.08943         \\ \hline
$^{91}$Rb   &{\bf\color{red}{~783.40752~(0.02400)}}     \cite{XianW2024PRC}      &{\bf\color{red}{~783.28815~(0.00780) }}   &~   0.11937         \\ \hline
$^{99}$Rb   &~821.21648~(0.03100)     \cite{Mukul2021PRC}                        &~821.23480~(0.00403)                      &~  $-$0.01832         \\ \hline
$^{100}$Rb  &~824.42979~(0.03000)     \cite{Mukul2021PRC}                        &~824.45085~(0.01312)                      &~  $-$0.02106         \\ \hline
$^{101}$Rb  &{\bf\color{red}{~828.73810~(0.02900) }}    \cite{Mukul2021PRC}      &{\bf\color{red}{~828.82371~(0.02049)}}    &~  $-$0.08561         \\ \hline
$^{102}$Rb  &~831.57041~(0.02900)     \cite{Mukul2021PRC}                        &~831.57992~(0.08290)                      &~  $-$0.00951         \\ \hline
$^{103}$Rb  &~835.44972~(0.03200)     \cite{Mukul2021PRC}                        &~$^{\#}$835.53600~(0.00400)               &~  $-$0.08628         \\ \hline
$^{75}$Sr   &~{\bf\color{red}{~621.82176~(0.15000)}}     \cite{wang2023mass}     &{\bf\color{red}{~622.24176~(0.22000)}}    &~  $-$0.41661         \\ \hline
$^{79}$Sr   &~673.38170~(0.02600)     \cite{xing2023isochronous}                 &~673.38461~(0.00742)                      &~  $-$0.00291         \\ \hline
$^{99}$Sr   &~831.84220~(0.03100)     \cite{Mukul2021PRC}                        &~831.84983~(0.00474)                      &~  $-$0.00763         \\ \hline
$^{100}$Sr  &~837.22851~(0.02900)     \cite{Mukul2021PRC}                        &~837.22012~(0.00692)                      &~   0.00839         \\ \hline
$^{101}$Sr  &~840.78682~(0.02900)     \cite{Mukul2021PRC}                        &~840.79885~(0.00848)                      &~  $-$0.01203         \\ \hline
$^{102}$Sr  &~845.72213~(0.02900)     \cite{Mukul2021PRC}                        &~845.70457~(0.06707)                      &~   0.01756         \\ \hline
$^{103}$Sr  &~848.83844~(0.02900)     \cite{Mukul2021PRC}                        &~$^{\#}$848.92600~(0.20000)               &~  $-$0.08756         \\ \hline
$^{104}$Sr  &~853.10075~(0.03300)     \cite{Mukul2021PRC}                        &~$^{\#}$853.42400~(0.30000)               &~  $-$0.32325         \\ \hline
$^{105}$Sr  &~855.64706~(0.04400)     \cite{Mukul2021PRC}                        &~$^{\#}$855.96000~(0.50000)               &~  $-$0.31294         \\ \hline
$^{81}$Y    &~688.98805~(0.02800) \cite{xing2023isochronous}                     &~688.98025~(0.00541)                      &~   0.00989         \\ \hline
$^{104}$Y   &~863.01896~(0.01600) \cite{Hukkanen2024PLB}                         &~$^{\#}$862.99200~(0.20000)               &~   0.02696         \\ \hline
$^{80}$Zr   &~669.54176~(0.02000) \cite{Hamaker2021NaturePhys}                   &~$^{\#}$669.20000~(0.30000)               &~   0.34176         \\ \hline
$^{81}$Zr   &~680.04107~(0.01000) \cite{Hamaker2021NaturePhys}                   &~680.00730~(0.09222)                      &~   0.03377         \\ \hline
$^{82}$Zr   &{\bf\color{red}{~694.17498~(0.00250)}} \cite{Hamaker2021NaturePhys} &{\bf\color{red}{~694.16826~(0.00158)}}    &~   0.00672         \\ \hline
$^{83}$Zr   &~704.54402~(0.00640) \cite{Hamaker2021NaturePhys}                   &~704.53718~(0.00643)                      &~   0.00684         \\ \hline
$^{106}$Zr  &~882.85052~(0.04300) \cite{Hukkanen2024PLB}                         &~$^{\#}$882.98000~(0.20000)               &~  $-$0.01076         \\ \hline
$^{104}$Nb  &{\bf\color{red}{~879.15532~(0.01800)}} \cite{Hukkanen2024PLB}       &{\bf\color{red}{~879.15185~(0.00178)}}    &~   0.11995         \\ \hline
$^{109}$Nb  &~904.63155~(0.18000) \cite{Hukkanen2024PLB}                         &~904.38725~(0.43082)                      &~   0.24430         \\ \hline
$^{87}$Mo   &~736.23138~(0.03300) \cite{xing2023isochronous}                     &~736.23091~(0.00286)                      &~   0.00046         \\ \hline
$^{94}$Mo   &~814.26160~(0.13600) \cite{Horana2022PRC}                           &~814.25941~(0.00014)                      &~   0.00224         \\ \hline
$^{111}$Mo  &~922.99982~(0.01100) \cite{Hou2023PRC}                              &~922.99755~(0.01258)                      &~   0.00227         \\ \hline
$^{112}$Mo  &~928.70557~(0.11800) \cite{Hukkanen2024PLB}                         &~$^{\#}$928.61113~(0.20000)               &~   0.11357         \\ \hline
$^{115}$Tc  &~949.79058~(0.87400) \cite{WangKL2024PRC}                           &$^{\#}$950.36000~(0.00200)                &~  $-$0.56942         \\ \hline
$^{91}$Ru   &~768.31007~(0.05900) \cite{xing2023isochronous}                     &~768.30651~(0.00222)                      &~   0.00355         \\ \hline
$^{108}$Ru  &{\bf\color{red}{~921.06220~(0.01200)}} \cite{PorterPRC110}          &{\bf\color{red}{~920.94048~(0.00868)}}    &~   0.12172         \\ \hline
$^{110}$Ru  &{\bf\color{red}{~933.61999~(0.07700)}} \cite{PorterPRC110}          &{\bf\color{red}{~933.49435~(0.00892)}}    &~   0.12564         \\ \hline
$^{111}$Ru  &~938.28427~(0.01200) \cite{Hou2023PRC}                              &~938.27836~(0.00968)                      &~   0.00591         \\ \hline
$^{112}$Ru  &~945.19358~(0.00800) \cite{Hou2023PRC}                              &~945.19523~(0.00960)                      &~  $-$0.00165         \\ \hline
$^{113}$Ru  &~949.50389~(0.00700) \cite{Hou2023PRC}                              &~949.50348~(0.03828)                      &~   0.00040         \\ \hline
$^{93}$Rh   &~784.43541~(0.02000) \cite{xing2023isochronous}                     &~784.43877~(0.00263)                      &~  $-$0.00336         \\ \hline
$^{111}$Rh  &~943.02999~(0.01300) \cite{Hou2023PRC}                              &~943.01456~(0.00685)                      &~   0.01543         \\ \hline
$^{112}$Rh  &{\bf\color{red}{~948.36330~(0.01800)}} \cite{Hou2023PRC}            &{\bf\color{red}{~948.51306~(0.04409)}}    &~  $-$0.14976         \\ \hline
$^{113}$Rh  &{\bf\color{red}{~955.63861~(0.01600)}} \cite{Hou2023PRC}            &{\bf\color{red}{~955.62026~(0.00713)}}    &~   0.01835         \\ \hline
$^{95}$Pd   &~800.75675~(0.02000) \cite{xing2023isochronous}                     &~800.75317~(0.00303)                      &~   0.00359         \\ \hline
$^{111}$Pd  &{\bf\color{red}{~945.94471~(0.01100)}} \cite{Hou2023PRC}            &{\bf\color{red}{~945.91423~(0.00073)}}    &~   0.03049         \\ \hline
$^{112}$Pd  &~954.32502~(0.02400) \cite{Hou2023PRC}                              &~954.32302~(0.00655)                      &~   0.00433         \\ \hline
$^{113}$Pd  &~959.64833~(0.04800) \cite{Hou2023PRC}                              &~959.66433~(0.00695)                      &~  $-$0.01315         \\ \hline
$^{116}$Pd  &{\bf\color{red}{~980.25022~(0.08100)}} \cite{PorterPRC110}          &{\bf\color{red}{~980.11596~(0.00714)}}    &~   0.13426         \\ \hline
$^{123}$Pd  &~1017.06843~(0.26500)\cite{LiHF2022PRL}                             &~1017.21391~(0.78944)                     &~  $-$0.14548         \\ \hline
$^{95}$Ag   &~789.85718~(0.01400) \cite{GeZh2024PRL}                             &~$^{\#}$789.92500~(0.40000)               &~  $-$0.06782         \\ \hline
$^{96}$Ag   &{\bf\color{red}{~802.84300~(0.09500)}} \cite{GeZh2024PRL}           &{\bf\color{red}{~802.58787~(0.09008)}}    &~   0.25513         \\ \hline
$^{97}$Ag   &{\bf\color{red}{~817.19404~(0.01400)}} \cite{GeZh2024PRL}           &{\bf\color{red}{~817.05157~(0.01202)}}    &~   0.14247         \\ \hline
$^{111}$Ag  &{\bf\color{red}{~947.34244~(0.01700) }} \cite{Hou2023PRC}           &{\bf\color{red}{~947.36344~(0.00146)}}    &~  $-$0.01901         \\ \hline
$^{113}$Ag  &~962.28606~(0.04100) \cite{Hou2023PRC}                              &~962.31806~(0.01664)                      &~  $-$0.02940         \\ \hline
$^{118}$Cd  &~1001.55533~(0.02000) \cite{JariesPRC2023}                          &~1001.56693~(0.02000)                     &~  $-$0.00920         \\ \hline
$^{119}$Cd  &{\bf\color{red}{~1007.00144~(0.02100)  }}  \cite{JariesPRC2023}     &{\bf\color{red}{~1006.91354~(0.03770)}}   &~   0.09031         \\ \hline
$^{125}$Cd  &~1044.60150~(0.32000)\cite{LiHF2022PRL}                             &1044.71019(0.00289)                       &~  $-$0.10869         \\ \hline
$^{99}$In   &~822.15717~(0.07700) \cite{mougeot2021mass}                         &~$^{\#}$822.10817~(0.29800)               &~   0.06117         \\ \hline
$^{100}$In  &~832.98648~(0.02000) \cite{mougeot2021mass}                         &~832.97758~(0.00224)                      &~   0.01153         \\ \hline
$^{101}$In  &~845.41619~(0.04700) \cite{mougeot2021mass}                         &~845.41579~(0.01166)                      &~   0.00316         \\ \hline
$^{117}$In  &~994.95465~(0.02400) \cite{JariesPRC2023}                           &~994.95475~(0.00488)                      &~   0.00241         \\ \hline
$^{118}$In  &{\bf\color{red}{~1001.36016~(0.02000)}} \cite{JariesPRC2023}        &{\bf\color{red}{~1001.30871~(0.00775)}}   &~   0.05145         \\ \hline
$^{119}$In  &~1009.86657~(0.02100) \cite{JariesPRC2023}                          &~1009.85050~(0.00731)                     &~   0.01606         \\ \hline
$^{120}$In  &~1015.93478~(0.03100) \cite{Nesterenko2023PRC}                      &~1015.95090~(0.04001)                     &~  $-$0.01612         \\ \hline
$^{121}$In  &~1024.14199~(0.01200) \cite{Nesterenko2023PRC}                      &~1024.12908~(0.02742)                     &~   0.01291         \\ \hline
$^{123}$In  &~1037.83821~(0.01100) \cite{Nesterenko2023PRC}                      &~1037.86615~(0.01983)                     &~  $-$0.02794         \\ \hline
$^{124}$In  &~1043.42142~(0.03200) \cite{Nesterenko2023PRC}                      &~1043.37621~(0.03056)                     &~   0.04521         \\ \hline
$^{126}$In  &~1056.36054~(0.26900) \cite{LiHF2022PRL}                            &~1056.46044~(0.00419)                     &~  $-$0.09990         \\ \hline
$^{127}$In  &~1063.59785~(0.03700) \cite{Izzo2021PRC}                            &~1063.60229~(0.01000)                     &~  $-$0.00444         \\ \hline
$^{128}$In  &~1068.97916~(0.03800) \cite{Izzo2021PRC}                            &~1068.98382~(0.00132)                     &~  $-$0.00466         \\ \hline
$^{129}$In  &~1075.71347~(0.03700) \cite{Izzo2021PRC}                            &~1075.69991~(0.00197)                     &~   0.01356         \\ \hline
$^{131}$In  &~1087.06109~(0.04000) \cite{Izzo2021PRC}                            &~1087.03203~(0.00221)                     &~   0.02906         \\ \hline
$^{132}$In  &~1089.47640~(0.03800) \cite{Izzo2021PRC}                            &~1089.49054~(0.06003)                     &~  $-$0.01414         \\ \hline
$^{133}$In  &~1092.83071~(0.04100) \cite{Izzo2021PRC}                            &~$^{\#}$1092.86100~(0.20000)              &~  $-$0.03029         \\ \hline
$^{134}$In  &~1095.07902~(0.04400) \cite{Izzo2021PRC}                            &~$^{\#}$1095.18200~(0.20000)              &~  $-$0.10298         \\ \hline
$^{103}$Sn  &~859.36913~(0.06800)  \cite{xing2023isochronous}                    &~$^{\#}$859.32313~(0.10000)               &~   0.04013         \\ \hline
$^{104}$Sb  &~ 858.86016~(0.07000) \cite{xing2023isochronous}                    &~$^{\#}$858.83200~(0.10100)               &~   0.02816         \\ \hline
$^{128}$Sb  &~1077.86160~(0.00210) \cite{Hoff2023PRL}                            &~1077.85887~(0.01879)                     &~   0.00273         \\ \hline
$^{133}$Sb  &{\bf\color{red}{~1112.65801~(0.01100)}} \cite{Valverde2024PLB}      &{\bf\color{red}{~1112.50912~(0.00313)}}   &~   0.14889         \\ \hline
$^{137}$Sb  &~1125.88469~(0.06300) \cite{Beliuskina2024PRC}                      &~1125.93127~(0.05216)                     &~  $-$0.04658         \\ \hline
$^{107}$Te  &~883.65182~(0.07000)  \cite{xing2023isochronous}                    &~$^{\#}$883.60600~(0.10100)               &~   0.04582         \\ \hline
$^{133}$Te  &{\bf\color{red}{~1115.73988~(0.00700)}} \cite{Valverde2024PLB}      &{\bf\color{red}{~1115.74039~(0.00207)}}   &~   0.14845         \\ \hline
$^{108}$I   &~ 883.18181~(0.07000) \cite{xing2023isochronous}                    &~$^{\#}$883.00800~(0.10100)               &~   0.17381         \\ \hline
$^{133}$I   &{\bf\color{red}{~1118.04126~(0.00900)}} \cite{Valverde2024PLB}	     &{\bf\color{red}{~1117.87821~(0.00590)}}   &~   0.16305         \\ \hline
$^{136}$I   &~1135.78803~(0.04400) \cite{Beliuskina2024PRC}                      &~1135.78009~(0.01419)                     &~   0.00794         \\ \hline
$^{137}$I   &~1140.67091~(0.04800) \cite{Beliuskina2024PRC}                      &~1140.66245~(0.00838)                     &~   0.00846         \\ \hline
$^{138}$I   &~1144.34349~(0.08200) \cite{Beliuskina2024PRC}                      &~1144.35741~(0.00596)                     &~  $-$0.01392         \\ \hline
$^{139}$I   &~1148.92106~(0.02200) \cite{Beliuskina2024PRC}                      &~1148.91979~(0.00401)                     &~   0.00127         \\ \hline
$^{141}$I   &~1156.50528~(0.04500) \cite{Beliuskina2024PRC}                      &~1156.51812~(0.01584)                     &~  $-$0.01284         \\ \hline
$^{142}$I   &~1159.50839~(0.05900) \cite{Beliuskina2024PRC}                      &~1159.46574~(0.00494)                     &~   0.04265         \\ \hline
$^{111}$Xe  &~908.24251 ~(0.09000) \cite{xing2023isochronous}                    &$^{\#}$908.20200~(0.11500)                &~   0.04051         \\ \hline
$^{112}$Cs  &~907.42154 ~(0.09000) \cite{xing2023isochronous}                    &$^{\#}$907.42400~(0.11600)                &~  $-$0.00246         \\ \hline
$^{151}$La  &~1227.88420~(0.02600) \cite{Kimura2024PRC}                          &~1227.48558~(0.43547)                     &~   0.39862         \\ \hline
$^{151}$Ce  &{\bf\color{red}{~1234.78792~(0.02400)}} \cite{Kimura2024PRC}        &{\bf\color{red}{~1234.61796~(0.01770)}}   &~   0.16996         \\ \hline
$^{152}$Ce  &~1240.52835~(0.02000) \cite{Kimura2024PRC}                          &~$^{\#}$1240.61735~(0.20000)              &~   0.05635         \\ \hline
$^{153}$Ce  &~1244.42778~(0.03300) \cite{Kimura2024PRC}                          &~$^{\#}$1244.61978~(0.20000)              &~  $-$0.07422         \\ \hline
$^{154}$Ce  &~1249.67863~(0.02400) \cite{Orford2022PRC}                          &~$^{\#}$1249.86400~(0.20000)              &~  $-$0.18537         \\ \hline
$^{152}$Pr  &~1244.46704~(0.09800) \cite{Kimura2024PRC}                          &~1244.43995~(0.01854)                     &~   0.02709         \\ \hline
$^{153}$Pr  &{\bf\color{red}{~1250.48351~(0.03300)}} \cite{Kimura2024PRC}        &{\bf\color{red}{~1250.32168~(0.01188)}}   &~   0.16183         \\ \hline
$^{154}$Pr  &{\bf\color{red}{~1254.82816~(0.02500)}} \cite{Orford2022PRC}        &{\bf\color{red}{~1254.68411~(0.10001)}}   &~   0.14405         \\ \hline
$^{156}$Pr  &~1264.41928~(0.01000) \cite{Orford2022PRC}                          &~1264.41646~(0.00103)                     &~   0.00282         \\ \hline
$^{157}$Pr  &~1269.47609~(0.03200) \cite{Orford2022PRC}                          &~1269.47327~(0.00317)                     &~   0.00282         \\ \hline
$^{151}$Nd  &{\bf\color{red}{~1242.94637~(0.02300)}} \cite{Kimura2024PRC}        &{\bf\color{red}{~1242.77139~(0.00113)}}   &~   0.17498         \\ \hline
$^{152}$Nd  &{\bf\color{red}{~1250.22080~(0.02000)}} \cite{Kimura2024PRC}        &{\bf\color{red}{~1250.04919~(0.02448)}}   &~   0.17161         \\ \hline
$^{157}$Nd  &~1276.75331~(0.02200) \cite{Orford2022PRC}                          &~1276.75024~(0.00214)                     &~   0.00308         \\ \hline
$^{151}$Pm  &{\bf\color{red}{~1244.60409~(0.02400)}} \cite{Kimura2024PRC}        &{\bf\color{red}{~1244.43212~(0.00461)}}   &~   0.17197         \\ \hline
$^{152}$Pm  &{\bf\color{red}{~1250.50752~(0.02000)}} \cite{Kimura2024PRC}        &{\bf\color{red}{~1250.37165~(0.02590)}}   &~   0.13587         \\ \hline
$^{153}$Pm  &{\bf\color{red}{~1258.04495~(0.03300)}} \cite{Kimura2024PRC}        &{\bf\color{red}{~1257.83650~(0.00906)}}   &~   0.20845         \\ \hline
$^{161}$Pm  &~1301.84857~(0.09200) \cite{Orford2022PRC}                          &~1301.84563~(0.00904)                     &~   0.00294         \\ \hline
$^{153}$Sm  &{\bf\color{red}{~1259.14868~(0.03300)}} \cite{Kimura2024PRC}        &{\bf\color{red}{~1258.96621~(0.00103)}}   &~   0.18247         \\ \hline
$^{163}$Eu  &~1322.91397~(0.09000) \cite{Orford2022PRC}                          &~1322.91080~(0.00090)                     &~   0.00317         \\ \hline
$^{165}$Eu  &~1333.21776~(0.01000) \cite{Orford2022PRC}                          &~1333.20873~(0.00521)                     &~   0.00903         \\ \hline
$^{148}$Tb  &~1214.24094~(0.01200) \cite{lykiardopoulou2023exploring}            &~1214.23946~(0.01246)                     &~   0.00148         \\ \hline
$^{155}$Tb  &~1271.46001~(0.01200) \cite{Orford2022PRC}                          &~1271.45218~(0.00983)                     &~   0.00783         \\ \hline
$^{152}$Ho  &~1238.02963~(0.01200) \cite{lykiardopoulou2023exploring}            &~1238.02773~(0.01253)                     &~   0.00189         \\ \hline
$^{151}$Er  &~1223.84104~(0.01500) \cite{lykiardopoulou2023exploring}            &~1223.83572~(0.01647)                     &~   0.00533         \\ \hline
$^{152}$Tm  &~1224.55708~(0.05100) \cite{lykiardopoulou2023exploring}            &~1224.57868~(0.05403)                     &~  $-$0.02161         \\ \hline
$^{156}$Tm  &~1261.97832~(0.01400) \cite{lykiardopoulou2023exploring}            &~1261.97810~(0.01428)                     &~   0.00022         \\ \hline
$^{157}$Tm  &~1271.93463~(0.02400) \cite{lykiardopoulou2023exploring}            &~1271.92428~(0.02795)                     &~   0.01035         \\ \hline
$^{150}$Yb  &~1194.57218~(0.04500) \cite{Beck2021PRL}                            &~$^{\#}$1194.76718~(0.30000)              &~  $-$0.17782         \\ \hline
$^{151}$Yb  &~1205.33449~(0.10600) \cite{Beck2021PRL}                            &~1205.55049~(0.30049)                     &~  $-$0.21261         \\ \hline
$^{152}$Yb  &~1218.15880~(0.04400) \cite{Beck2021PRL}                            &~1218.34980~(0.14971)                     &~  $-$0.18764         \\ \hline
$^{153}$Yb  &~1227.25311~(0.04600) \cite{Beck2021PRL}                            &~$^{\#}$1227.31111~(0.20000)              &~  $-$0.11289         \\ \hline
$^{154}$Yb  &~1238.15642~(0.04500) \cite{Beck2021PRL}                            &~1238.15442~(0.01728)                     &~   0.00563         \\ \hline
$^{155}$Yb  &~1246.79873~(0.01600) \cite{Beck2021PRL}                            &~1246.79673~(0.01660)                     &~   0.00610         \\ \hline
$^{156}$Yb  &{\bf\color{red}{~1257.69604~(0.05500)}} \cite{Beck2021PRL}          &{\bf\color{red}{~1257.63104~(0.00931)}}   &~   0.06917         \\ \hline
$^{157}$Yb  &~1265.83135~(0.05400) \cite{Beck2021PRL}                            &~1265.85635~(0.01091)                     &~  $-$0.02120         \\ \hline
$^{151}$Lu  &~1193.33121~(0.04500) \cite{lykiardopoulou2023exploring}            &~$^{\#}$1193.52621~(0.30000)              &~  $-$0.17279         \\ \hline
$^{153}$Lu  &~1217.55283~(0.04500) \cite{lykiardopoulou2023exploring}            &~1217.74383~(0.15002)                     &~  $-$0.18766         \\ \hline
$^{154}$Lu  &~1227.04914~(0.04800) \cite{lykiardopoulou2023exploring}            &~$^{\#}$1227.10714~(0.20100)              &~  $-$0.02286         \\ \hline
$^{156}$Lu  &~1247.25776~(0.05100) \cite{lykiardopoulou2023exploring}            &~1247.28276~(0.05412)                     &~  $-$0.02077         \\ \hline
$^{156}$Hf  &~1240.42849~(0.04400) \cite{lykiardopoulou2023exploring}            &~1240.62049~(0.14974)                     &~  $-$0.18766         \\ \hline
$^{157}$Hf  &~1249.66880~(0.04600) \cite{lykiardopoulou2023exploring}            &~$^{\#}$1249.72680~(0.20000)              &~  $-$0.05120         \\ \hline
$^{159}$Hf  &~1269.86942~(0.01600) \cite{lykiardopoulou2023exploring}            &~1269.86742~(0.01681)                     &~   0.00622         \\ \hline
$^{157}$Ta  &~1239.49352~(0.04500) \cite{lykiardopoulou2023exploring}            &~1239.68552~(0.15005)                     &~  $-$0.18802         \\ \hline
$^{158}$Ta  &~1249.22183~(0.04800)  \cite{lykiardopoulou2023exploring}           &~$^{\#}$1249.27883~(0.20100)              &~  $-$0.08417         \\ \hline
$^{160}$Ta  &~1270.10245~(0.05100)  \cite{lykiardopoulou2023exploring}           &~1270.12745~(0.05432)                     &~  $-$0.02081         \\ \hline
$^{160}$W   &~1262.65818~(0.04400)  \cite{lykiardopoulou2023exploring}           &~1262.85018~(0.14981)                     &~  $-$0.18811         \\ \hline
$^{161}$W   &~1272.04149~(0.04600)  \cite{lykiardopoulou2023exploring}           &~$^{\#}$1272.09949~(0.20000)              &~  $-$0.01951         \\ \hline
$^{163}$W   &~1292.64511~(0.05800)  \cite{lykiardopoulou2023exploring}           &~1292.64311~(0.05843)                     &~   0.00550         \\ \hline
$^{161}$Re  &~1261.46121~(0.04400)  \cite{lykiardopoulou2023exploring}           &~1261.65321~(0.14991)                     &~  $-$0.18779         \\ \hline
$^{162}$Re  &~1271.27652~(0.04700)  \cite{lykiardopoulou2023exploring}           &~$^{\#}$1271.33452~(0.20100)              &~  $-$0.09948         \\ \hline
$^{164}$Re  &~1292.47114~(0.05100)  \cite{lykiardopoulou2023exploring}           &~1292.49614~(0.05456)                     &~ $-$0.02144         \\ \hline
$^{164}$Os  &~1284.47486~(0.04400)  \cite{lykiardopoulou2023exploring}           &~1284.66686~(0.14990)                     &~  $-$0.18765         \\ \hline
$^{165}$Os  &~1294.00217~(0.04600)  \cite{lykiardopoulou2023exploring}           &~$^{\#}$1294.06017~(0.20000)              &~  $-$0.09283         \\ \hline
$^{167}$Os  &~1314.95679~(0.08100)  \cite{lykiardopoulou2023exploring}           &~1314.95479~(0.08089)                     &~   0.00619         \\ \hline
$^{165}$Ir  &~$^{\#}$1282.93390~(0.06700) \cite{lykiardopoulou2023exploring}   &~$^{\#}$1283.12590~(0.15800)              &~  $-$0.10610         \\ \hline
$^{166}$Ir  &~1292.85021~(0.04700)  \cite{lykiardopoulou2023exploring}           &~$^{\#}$1292.90821~(0.20100)              &~  $-$0.12379         \\ \hline
$^{168}$Ir  &~1314.38583~(0.05200)  \cite{lykiardopoulou2023exploring}           &~1314.41083~(0.05522)                     &~  $-$0.02112         \\ \hline
$^{168}$Pt  &~1305.78055~(0.04400)  \cite{lykiardopoulou2023exploring}           &~1305.97255~(0.14993)                     &~  $-$0.18789         \\ \hline
$^{169}$Pt  &~1315.44086~(0.04700)  \cite{lykiardopoulou2023exploring}           &~$^{\#}$1324.49786~(0.20000)              &~  $-$0.05514         \\ \hline
$^{171}$Pt  &~1336.64548~(0.08100)  \cite{lykiardopoulou2023exploring}           &~1336.64348~(0.08095)                     &~   0.00659         \\ \hline
$^{170}$Au  &~1313.96890~(0.04800)  \cite{lykiardopoulou2023exploring}           &~$^{\#}$1314.02590~(0.20100)              &~  $-$0.13110         \\ \hline
$^{172}$Au  &~1335.75852~(0.05300)  \cite{lykiardopoulou2023exploring}           &~1335.78352~(0.05616)                     &~  $-$0.02078         \\ \hline
$^{172}$Hg  &~1326.55224~(0.04500)  \cite{lykiardopoulou2023exploring}           &~1326.74424~(0.15006)                     &~  $-$0.18806         \\ \hline
$^{173}$Hg  &~1336.35855~(0.04700)  \cite{lykiardopoulou2023exploring}           &~$^{\#}$1336.41555~(0.20100)              &~  $-$0.06645         \\ \hline
$^{175}$Hg  &~1357.86817~(0.08100)  \cite{lykiardopoulou2023exploring}           &~1357.86617~(0.08009)                     &~   0.00582         \\ \hline
$^{176}$Tl  &~1357.76920~(0.08300)  \cite{lykiardopoulou2023exploring}           &~1356.60120~(0.08306)                     &~   0.00606         \\ \hline
$^{235}$Pa  &~1783.30073~(0.04400)  \cite{Niwase2023PRL}                         &~1783.27727~(0.01397)                     &~   0.02346         \\ \hline
$^{236}$Pa  &~1788.30104~(0.07400)  \cite{Niwase2023PRL}                         &~1788.30352~(0.01397)                     &~  $-$0.00248         \\ \hline
$^{237}$Pa  &~1794.14935~(0.04400)  \cite{Niwase2023PRL}                         &~1794.18117~(0.01304)                     &~  $-$0.03182         \\ \hline
$^{235}$U   &~1783.89546~(0.04400)  \cite{Niwase2023PRL}                         &~1783.86505~(0.00112)                     &~   0.03041         \\ \hline
$^{236}$U   &~1790.44777~(0.07400)  \cite{Niwase2023PRL}                         &~1790.41055~(0.00111)                     &~   0.03722         \\ \hline
$^{237}$U   &{\bf\color{red}{~1795.58308~(0.04200)}}  \cite{Niwase2023PRL}       &{\bf\color{red}{~1795.53632~(0.00120)}}   &~   0.04676         \\ \hline
$^{238}$U   &{\bf\color{red}{~1801.71339~(0.02000)}}  \cite{Niwase2023PRL}       &{\bf\color{red}{~1801.69004~(0.00149)}}   &~   0.02335         \\ \hline
$^{239}$U   &~1806.48870~(0.02000)  \cite{Niwase2023PRL}                         &~1806.49641~(0.00150)                     &~  $-$0.00771         \\ \hline
$^{240}$U   &~1812.41201~(0.01600)  \cite{Niwase2023PRL}                         &~1812.42492~(0.00255)                     &~  $-$0.01291         \\ \hline
$^{241}$U   &~1817.03432~(0.04500)  \cite{Niwase2023PRL}                         &~$^{\#}$1816.89900~(0.19600)              &~   0.13532         \\ \hline
$^{242}$U   &~1822.64363~(0.09000)  \cite{Niwase2023PRL}                         &~$^{\#}$1822.74400~(0.20100)              &~  $-$0.10037         \\ \hline
$^{239}$Np  &~1806.96942~(0.01900) \cite{Niwase2023PRL}                          &~1806.97575~(0.00131)                     &~  $-$0.00633         \\ \hline
$^{240}$Np  &~1812.04073~(0.02800) \cite{Niwase2023PRL}                          &~1812.04183~(0.01703)                     &~ $-$0.00110         \\ \hline
$^{241}$Np  &~1818.08204~(0.03100) \cite{Niwase2023PRL}                          &~1818.11427~(0.10001)                     &~  $-$0.03223         \\ \hline
$^{242}$Np  &~1822.99635~(0.08100) \cite{Niwase2023PRL}                          &~1823.08382~(0.20000)                     &~  $-$0.08747         \\ \hline
$^{239}$Pu  &~1806.90115~(0.04000) \cite{Niwase2023PRL}                          &~1806.91617~(0.00111)                     &~  $-$0.01502         \\ \hline
$^{240}$Pu  &{\bf\color{red}{~1813.48746~(0.02900)}} \cite{Niwase2023PRL}        &{\bf\color{red}{~1813.45039~(0.00111)}}   &~   0.03707         \\ \hline
$^{241}$Pu  &~1818.64377~(0.06100) \cite{Niwase2023PRL}                          &~1818.69192~(0.00111)                     &~  $-$0.04815         \\ \hline
$^{242}$Pu  &{\bf\color{red}{~1825.07908~(0.06600)}} \cite{Niwase2023PRL}        &{\bf\color{red}{~1825.00147~(0.00125)}}   &~   0.07761         \\ \hline
\end{longtable*}

\begin{table*}[h!]
\caption{Measurement methods and sources of the experimental data in TABLE 
\ref{mass-table}.}\label{exp-method}
\begin{tabularx}{\textwidth}{c|l|l}
\hline\hline
  {\bf Method}~ &~~  {\bf Source}        &~~{\bf References} \\ \hline
  B$\rho$-time-of-flight (B$\rho$-TOF)  &~ S800 spectrograph                       &~~\cite{WangKL2024PRC} \\ \hline
  B$\rho$-defined isochronous            &~ second Radiactive Isotope Beam Line in  & \\
mass spectrometry (B$\rho$-IMS)          &~ Lanzhou-experimental Cooler Storage Ring (RIBLL2-CSRe)   &~~\cite{YUY2024PRL133,WangM2022PRC,wang2023mass,zhou2023Mass} \\ \hline
  Storage ring                           &~ RIBLL2-CSRe                             &~~\cite{xing2023isochronous}  \\
  mass spectrometry                      &~ Radioactive Isotope Beam Factory (RIBF) &~~\cite{LiHF2022PRL}    \\ \hline
  \multirow{4}{*}{Ion trap}              &~ Isotope Separator On-Line Device tandem Penning trap mass&\\
                                         &~ spectrometer (ISOLTRAP)                  &~~\cite{mougeot2021mass}\\
                      &~ Canadian Penning trap (CPT)    &~~\cite{PorterPRC110,Hoff2023PRL,Valverde2024PLB,Orford2022PRC}  \\
                      &~ Low-Energy Beam and Ion Trap (LEBIT)    &~                                                      \cite{Puentes106PRC,Yandow2023PRC,Surbrook2021PRC,Porter2022PRC106,Lebit2021PRL,Horana2022PRC,Hamaker2021NaturePhys} \\
                      &~ Jyv\"askyl\"an Yliopiston Fysiikan Laitos (JYFLTRAP )  &~                            \cite{Canete2021PRC,giraud2022mass,Hukkanen2024PLB,GeZh2024PRL,JariesPRC2023,Nesterenko2023PRC,Beliuskina2024PRC} \\ \hline
                      &~ ISOLTRAP                                 &~~\cite{mougeot2021mass}\\
  Multiple-reflection &~ Fragment Separator Ion Catcher           &~~\cite{PhysRevC.103.034319} \\
  time-of-flight      &~ Gas Cell (GC)                            &~~\cite{Kimura2024PRC} \\
   (MR-TOF)           &~ KEK Isotope Separation System (KISS)     &~~\cite{Niwase2023PRL} \\
                      &~ RIBF                                     &~~\cite{Iimura2023PRL,XianW2024PRC,Hou2023PRC} \\
                      &~ TRIUMF's Ion Trap for Atomic and Nuclear science (TITAN)        &~  \cite{jacobs2023improved,Porter2022PRC106,Lebit2021PRL,Silwal2022PLB,Porter2022PRC105,PhysRevC.104.065803,Mukul2021PRC,Izzo2021PRC,lykiardopoulou2023exploring,Beck2021PRL} \\
\hline \hline
\end{tabularx}
\end{table*}

\end{appendix}

\end{document}